\documentclass[bibyear]{aa}  

\usepackage{natbib}
\usepackage{hyperref}
\hypersetup{colorlinks=true,linkcolor=[rgb]{1.,0.2,0.2},citecolor=[rgb]{0.1,0.4,1.},filecolor=[rgb]{0.7,0.2,0.2},urlcolor=[rgb]{0.7,0.2,0.2}}
\usepackage{xcolor}
\usepackage{orcidlink}
\usepackage{txfonts}

\usepackage{graphicx} 
\usepackage{color}    
\usepackage{float}
\usepackage{multicol}
\usepackage{amsmath}
\usepackage{verbatim}
\usepackage{array}
\usepackage{booktabs}
\usepackage{threeparttable}
\usepackage{amssymb}
\usepackage{array,multirow}
\usepackage{tikz}
\usetikzlibrary{shapes,backgrounds,calc,positioning,arrows.meta}

\newcommand{\stormy}{\emph{stormy}\xspace}
\newcommand{\cloudy}{\emph{cloudy}\xspace}
\newcommand{\rainy}{\emph{rainy}\xspace}
\newcommand{\sunny}{\emph{sunny}\xspace}

\begin{document} 

   \title{{\bfseries\scshape BlackHoleWeather –} Spin-coupled chaotic cold accretion across the meso scale: {\Large Kinematics and variability}}
    \titlerunning{SMBH spin evolution and jet-axis reorientation in chaotic cold accretion}
    \authorrunning{O.~Piana et al.}

\author{Olmo Piana \thanks{olmopiana@unimore.it}
        \inst{1}\orcidlink{0000-0002-1558-5289},
        Massimo Gaspari
        \inst{1}\orcidlink{0000-0003-2754-9258},
        Vieri Cammelli
        \inst{1}\orcidlink{0000-0002-2070-9047},
        Filippo Barbani
        \inst{1}\orcidlink{0000-0002-1620-2577},
        Davide M. Brustio
        \inst{1}\orcidlink{0009-0009-7700-1910},
        Giovanni Stel
        \inst{1}\orcidlink{0009-0007-0585-9462},
        Valeria Olivares
        \inst{2,3}\orcidlink{0000-0001-6638-4324},
        Ashkbiz Danehkar
        \inst{4}\orcidlink{0000-0003-4552-5997},
        Pasquale Temi
        \inst{5}\orcidlink{0000-0002-8341-342X},
        Roberto Serafinelli
        \inst{6,7}\orcidlink{0000-0003-1200-5071},
        Francesco Salvestrini
        \inst{8}\orcidlink{0000-0003-4751-7421},
        Filippo M. Maccagni
        \inst{9,10}\orcidlink{0000-0002-9930-1844},
        \and
        Francesco Tombesi
        \inst{6,11,12}\orcidlink{0000-0002-6562-8654}
        }

\institute{
Department of Physics, Informatics and Mathematics, University of Modena and Reggio Emilia, I-41125 Modena, Italy
\and
Department of Physics, Universidad de Santiago de Chile, Santiago, Chile
\and
CIRAS, Universidad de Santiago de Chile, Santiago, Chile
\and
Science and Technology Institute, Universities Space Research Association, Huntsville, AL 35805, USA
\and
NASA Ames Research Center, MS 245-6, Moffett Field, CA 94035-1000, USA
\and
INAF -- Astronomical Observatory of Rome, 00078 Monte Porzio Catone (Rome), Italy
\and
Instituto de Estudios Astrof\'isicos, Facultad de Ingenier\'ia y Ciencias, Universidad Diego Portales, Avenida Ej\'ercito Libertador 441, Santiago, Chile
\and
INAF -- Osservatorio Astronomico di Trieste, Via G. Tiepolo 11, I-34143 Trieste, Italy
\and
INAF -- Osservatorio Astronomico di Cagliari, via della Scienza 5, 09047, Selargius (CA), Italy
\and
Wits Centre for Astrophysics, School of Physics, University of the Witwatersrand, 2000, Johannesburg, South Africa
\and
Department of Physics, University of Rome ``Tor Vergata'', 00133 Rome, Italy
\and
INFN -- Rome ``Tor Vergata'' Section, 00133 Rome, Italy
}


\abstract
{Supermassive black hole (SMBH) spin records the vector history of accretion. In chaotic cold accretion (CCA), this history is set by clouds and filaments whose torques can add coherently, cancel, or reverse before reaching the horizon-scale closure.}
{We test whether halo stirring regulates SMBH spin by changing the radial continuity and torque coherence of the meso-scale accretion bridge. We focus on spin evolution, jet-axis reorientation, accretion variability, and CCA kinematics.}
{We analyse four 3D hydrodynamical simulations in a \(100\,{\rm kpc}\) box, reaching sub-pc resolution, including SMBH spin-coupled jet feedback. All runs use the \textit{Hybrid} SMBH spin model validated in a companion paper. Two simulations maintain continuous driven solenoidal turbulence, while two matched controls let the same initial turbulent field decay.}
{The main effect of persistent stirring is to disrupt mass and angular-momentum continuity across the meso-scale (pc--to--kpc) bridge. Although all runs develop comparable macro-scale inflow, in the driven-turbulence suite, gas struggles to reach pc scales, and the radial accretion rate drops by 2-3 orders of magnitude. Torque delivery in this case is fragmented and cancellation-dominated. The interrupted-turbulence suite, on the other hand, preserves a connected gas channel to the sink, while sustaining higher torque coherence. Driven runs therefore settle to slow effective jet-axis drift, whereas interrupted runs maintain reorientation rates higher by about two orders of magnitude and can briefly reach a few \({\rm deg\,Myr^{-1}}\) during coherent retrograde episodes. The same split appears in power spectra and k-plots: connected rain enhances low-frequency accretion power and produces narrower, phase-ordered kinematics, while stirring steepens high-frequency damping and broadens the gas velocity loci for all phases. All runs remain CCA-like, with turbulent Taylor numbers \(\mathrm{Ta_t}\lesssim1\).}
{}


\maketitle
%

\section{Introduction}
\label{sec:intro}

Supermassive black hole (SMBH) spin links unresolved angular-momentum transport to galaxy-scale jet power and orientation, and is therefore a key interface between accretion physics and feedback in the co-evolution between galaxies and their supermassive black holes \citep{kormendy2013}.
In the Blandford--Znajek picture \citep{blandford1977, tchekhovskoy2010}, the spin of the hole, together with the magnetic flux that threads the horizon, sets the efficiency with which rotational energy can be converted into jet power, while the angular momentum carried by the inflowing gas modifies both the magnitude and direction of the spin.
This two-way coupling offers one possible route to the diversity of observed jet and cavity morphologies, including misaligned X-ray bubbles and S- or Z-shaped radio structures \citep{krause2019, bruni2021, ubertosi2023}, although projection, jet-medium deflection, buoyant cavity evolution, and environmental asymmetry can also contribute. Such systems motivate models in which the jet axis can evolve, or be redirected, on tens and hundreds Myr timescales.

In this work, we study this coupling in the chaotic cold accretion regime \citep{gaspari2013, gaspari2015, gaspari2017}, the multiphase mode of SMBH feeding expected in hot halos subjected to radiative cooling, turbulence, and AGN heating \citep[see][for a review]{gaspari2020}. In CCA, radiative cooling and turbulent mixing precipitate warm and cold gas out of the hot atmosphere. The condensed gas then rains toward the centre through a network of clouds and filaments that continually reshuffle the angular-momentum budget of the inflow. CCA is therefore a demanding environment for any spin model, in which the SMBH is expected to be fed mostly by a stochastic sequence of three-dimensional torque episodes.
The central question of this paper is whether angular momentum retains enough coherence across the kpc-to-pc bridge (meso-scale) to reorient the SMBH spin and hence the jet axis efficiently.
In the previous paper \cite[][hereafter P26a]{piana2026a}, we introduced and validated an SMBH spin-evolution module coupled to the \textsc{AthenaPK} MHD code.
P26a established that, in our numerical setup, a general-relativistic ISCO-based closure for the angular momentum deposited onto the hole---the \textit{Hybrid} model---is required to avoid unrealistically large spin excursions, and that the model reproduces the expected behaviour in idealized prograde, retrograde, and Bardeen-type tests.
P26a also presented a first application to two turbulent CCA simulations, focusing mainly on their morphological and thermodynamic evolution. Here we extend that analysis with a matched interrupted-turbulence control suite, asking how the persistence of halo stirring regulates the meso-scale delivery of mass and angular momentum, and, in turn, the coherence of the torque acting on the SMBH. We connect the sink-scale dynamics to observables through torque coherence, effective jet-axis reorientation, accretion variability, and projected multiphase kinematics.

The present paper asks whether the feeding state of this meso-scale CCA flow leaves a measurable imprint on jet re-orientation, accretion variability, and the kinematic properties of the surrounding multiphase halo. We start then studying a multi-scale, physically consistent connection between feeding regimes, spin evolution, and jets, with the idea of informing the sub-grid models used in big-box cosmological simulations \citep{dubois2014, cielo2018, beckmann2019, bustamante2019, horton2020, talbot2021, talbot2022, beckmann2024} and semi-analytical models \citep{croton2006, bower2006, volonteri2008, hirschmann2012, barausse2012, sesana2014, mutch2016, lacey2016, marshall2019, piana2024, piana2025} which can be used to study population-wide statistics of SMBH spin across time. 

Within the \textsc{BlackHoleWeather} framework \citep[e.g.][]{gaspari2020}, and following the companion studies submitted, we use the CCA weather states as shorthand for how the meso-scale bridge connects halo rain to nuclear feeding. The non-jet simulations of \cite[][hereafter B26a,b]{barbani2026a, barbani2026b} isolate how turbulence shapes cold-rain topology, fragmentation, and sink delivery, while the fixed-axis jet simulations of \cite[][hereafter C26a,b]{cammelli2026a, cammelli2026b} show how feedback reshapes the same bridge and imprints observable multiphase diagnostics. The present paper, carried out within Work Package 3 (WP3) of the \textsc{BlackHoleWeather} project (PI: Gaspari), adds the spin-coupled vector layer: whether the bridge preserves enough radial continuity and directional memory to deliver coherent angular momentum to the SMBH.

In this \textsc{BlackHoleWeather} terminology, \emph{stormy} weather denotes extended, turbulent, filamentary precipitation with bursty circulation; \emph{rainy} weather a compact cold/warm channel efficiently connected to the sink; \emph{cloudy} weather fragmented multiphase gas with inefficient nuclear delivery; and \emph{sunny} weather a hot-dominated or feedback-cleared central state with weak feeding. Because these states are scale dependent and can overlap, the key question is how the meso-scale feeding bridge connects to the sink, and whether it is able to deliver mass and angular momentum coherently and for prolonged periods of time.

The paper is organized as follows. Section~\ref{sec:setup} describes the numerical setup and the simulation suites of driven-turbulence (DT) and interrupted-turbulence (IT). Sections~\ref{sec:jet} and \ref{sec:spin} summarize the spin-dependent jet prescription, the \textit{Hybrid} spin model, and the angular-momentum bookkeeping. Section~\ref{sec:results} presents the results, following the sequence from turbulent stirring and nuclear feeding to torque coherence, effective jet-axis reorientation, accretion variability, and multiphase CCA diagnostics. Section~\ref{sec:discussion} discusses the physical interpretation, including the spin-regulated weather cycle, and Section~\ref{sec:comparison} places the results in the context of previous spin, jet re-orientation, and CCA studies. Section~\ref{sec:conclusions} summarizes the main conclusions.


\section{Numerical setup} \label{sec:methods}

\subsection{Simulation setup and turbulent CCA suite}
\label{sec:setup}
The simulations analyzed here use the same numerical framework described in P26a, to which we refer the reader for the full methodological description. We summarize only the ingredients needed to interpret the present controlled experiment, emphasizing which quantities are held fixed and which are varied between the continuously driven and interrupted-turbulence suites.

All runs are performed with \textsc{AthenaPK} \citep{grete2023}, a GPU-accelerated AMR MHD code built on \textsc{Athena++} \citep{stone2020} through the \textsc{Parthenon} \citep{grete2023} and \textsc{Kokkos} \citep{edwards2014} libraries. 
The computational domain is a cube with \(100\,{\rm kpc}\) on the side, a root grid of \(128^3\) cells, and nested static mesh refinement (SMR) towards the centre, with 10 levels and a maximum spatial resolution \(\Delta x_{\rm min}=0.7\,{\rm pc}\) within the nuclear refinement region, \(r\lesssim22.5\,{\rm pc}\). A sink particle with radius $r_\mathrm{sink} \approx 3$ pc is located in the center of the computational domain.
The gravitational potential, entropy profile, radiative cooling, sink treatment, and jet injection algorithms are the same as in P26a, while the time evolution of the thermodynamic profiles is shown in Appendix \ref{app:profiles}. While the main details are reported in the next section, we refer the reader to B26a and C26a for the feeding and feedback modules, respectively, and to P26a for the numerical setup implementation.

\begin{figure}
\includegraphics[width=\columnwidth]{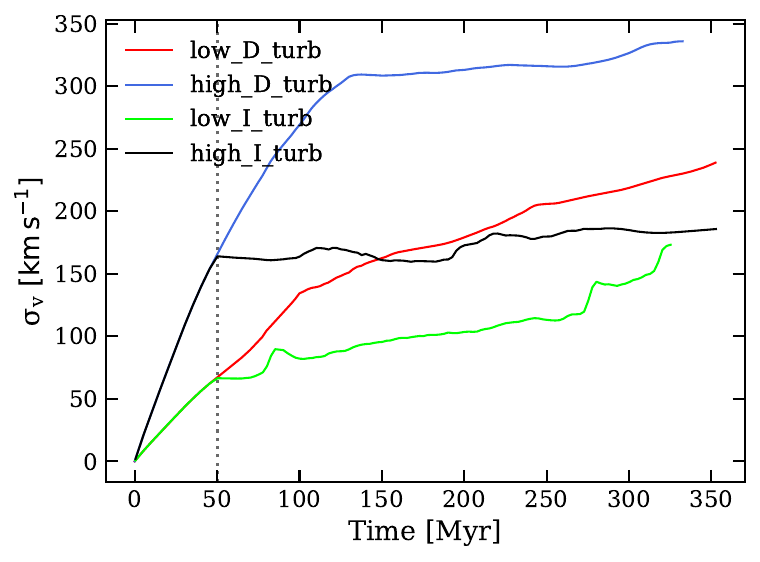}
\caption{Evolution of the volume-weighted velocity dispersion for the four runs. The average is performed across all phases and within the $100$-kpc box. We mask out the jet by imposing a velocity cut at $v = 1000\,{\rm km\,s^{-1}}$. The first $50\,{\rm Myr}$, indicated by the grey vertical line, correspond to the common pre-conditioning phase; the subsequent divergence quantifies whether stirring persists (DT) or is allowed to decay (IT).}
\label{sigmav}
\end{figure}

We analyze four simulations designed to study the effect of persistent halo stirring. Two are the continuously driven turbulence runs \texttt{low\_D\_turb} and \texttt{high\_D\_turb} already introduced in P26a, although in that case they were tagged as \texttt{hyb\_lowT\_s01} and \texttt{hyb\_highT\_s01}. The other two runs, \texttt{low\_I\_turb} and \texttt{high\_I\_turb}, are new and define the interrupted-turbulence suite. In both IT and DT runs, turbulence is driven for the first $50\,{\rm Myr}$, while radiative cooling and jet feedback are disabled. This stage imprints a volume-weighted velocity dispersion of $\sigma_v \approx 60\,{\rm km\,s^{-1}}$ for the \texttt{low\_D\_turb} and \texttt{low\_I\_turb} runs, and $\sigma_v \approx 160\,{\rm km\,s^{-1}}$ for the \texttt{high\_D\_turb} and \texttt{high\_I\_turb} runs. The complete evolution of the volume-weighted velocity dispersion is shown in Figure \ref{sigmav}. After $50\,{\rm Myr}$, the cooling and jets are switched on in all simulations; from that point onward, the turbulence continues to be driven in the DT runs but is turned off in the IT runs. Each DT/IT pair therefore starts from a matched entropy structure and a matched initial turbulent amplitude, so that the relevant difference during the feedback-regulated evolution is whether large-scale stirring persists. Subsonic turbulence is injected in solenoidal mode on scales of ${\sim}25\,{\rm kpc}$, and with a correlation time $t_{\rm corr}=30$ Myr \citep[e.g.][]{schmidt2009,grete2018,grete2025}. The initial SMBH spin is assumed to be $a_0 = 0.1$ and aligned with the z-axis, while its mass is $M_\mathrm{BH,0}= 2.8 \times 10^8 \mathrm{M_\odot}$ for all runs. See Table \ref{tab:sims} for details about the differences in parameters.

This distinction between continuously-driven and interrupted turbulence is the control parameter of the paper. The purpose of the DT and IT suites is to generate different feeding regimes at fixed thermodynamic initial conditions: one in which ongoing stirring persistently scrambles the inflow, and one in which the turbulent field is allowed to decay after the system enters the cooling--feedback loop.
A summary of the four runs is given in Table~\ref{tab:sims}. Because the morphology and thermodynamics of the driven runs were already discussed in P26a, the emphasis here is on diagnostics that track the path from feeding to observables: the duty cycle of the SMBH weather, the coherence of the delivered torque, the resulting jet-axis reorientation history, the CCA variability, and the multiphase kinematic signature of the halo via CCA diagnostics.

\begin{table*}[!ht]
\caption{Simulation suite analysed in this paper. The continuously driven runs are inherited from P26a, while the interrupted-turbulence runs are new.}
\label{tab:sims}
\centering
\begin{tabular}{lccccccc}
\toprule
Run & Driving & Box size [kpc] & $\Delta x_{\rm min}$ [pc] & $a_{\rm rms}$ [cm s$^{-2}$] & 3D Mach number & $\sigma_v (t_{50})$ [${\rm km\,s^{-1}}$] & Runtime [Myr]\\
\midrule
\texttt{low\_D\_turb}  & continuous & 100 & 0.7 & $6.2 \times 10^{-9}$ & $\sim 0.15$ & $60$--$70$ & $\approx 350$\\
\texttt{high\_D\_turb} & continuous & 100 & 0.7 & $1.55 \times 10^{-8}$ & $\sim 0.4$ & $160$--$170$ & $\approx 350$\\
\texttt{low\_I\_turb}  & interrupted & 100 & 0.7 & $6.2 \times 10^{-9}$ & $\sim 0.15$ & $60$--$70$ & $\approx 350$\\
\texttt{high\_I\_turb} & interrupted & 100 & 0.7 & $1.55 \times 10^{-8}$ & $\sim 0.4$ & $160$--$170$ & $\approx 350$\\
\bottomrule
\end{tabular}
\end{table*}

\subsection{Spin-jet coupling}
\label{sec:jet}

All simulations adopt the same jet prescription described in P26a, so we recall only the ingredients needed for the present interpretation. The jet is assumed to be powered through the Blandford--Znajek mechanism \citep{blandford1977}, so that the instantaneous feedback power depends on both the mass inflow reaching the innermost stable circular orbit (ISCO) and the current SMBH spin state:
\begin{equation}
  P_{\rm BZ} = \eta_{\rm BZ}\,\epsilon_{\rm isco}\,\dot{M}_{\rm sink}\,c^2,
\end{equation}
where $\dot{M}_{\rm sink}$ is the sink accretion rate, $\epsilon_{\rm isco}$ is the fraction of rest-mass energy available at the ISCO computed according to \cite{bardeen1972} and $\eta_{\rm BZ}$ is the spin-dependent jet efficiency. Specifically,
\begin{equation}
\epsilon_\mathrm{isco}=\frac{r_\mathrm{isco}^2 - 2r_\mathrm{isco} \pm a\sqrt{r_\mathrm{isco}}} {r_\mathrm{isco} \left(r_\mathrm{isco}^2 - 3r_\mathrm{isco} \pm 2a\sqrt{r_\mathrm{isco}}\right)^{1/2}},
\label{eq_eps_isco}
\end{equation}
where the dimensionless radius of the ISCO is calculated as $r_{\text{isco}} = 3 + Z_2 \mp \sqrt{(3 - Z_1)(3 + Z_1 + 2Z_2)}$, with the minus sign indicating prograde orbits and the plus sign retrograde orbits \citep{bardeen1972}. The auxiliary functions $Z_1$ and $Z_2$ are defined as in \citet{bardeen1972}.

As in \citet{EHT2019}, the Blandford--Znajek efficiency is written as
\begin{equation}
  \eta_{\rm BZ} = 2.8\,f(a)\left(\frac{\phi}{15}\right)^2,
\end{equation}
with
\begin{equation}
  f(a) = a^2\left(1+\sqrt{1-a^2}\right)^{-2},
\end{equation}
where \(a\) is the dimensionless spin magnitude and \(\phi=15\) is the magnetic-flux parameter, kept fixed in all runs. With this choice, variations in \(P_{\rm BZ}\) arise from the resolved sink accretion rate and from the spin-dependent factors \(\eta_{\rm BZ}(a)\) and \(\epsilon_{\rm isco}(a)\), while magnetic-flux evolution is held fixed. The prescription therefore isolates accretion- and spin-driven jet variability at fixed magnetic-flux normalization. 
A fully consistent magnetic field treatment is the object of future work.

The jet power is injected into two polar cylindrical regions at a distance of $\approx 5$ pc from the origin, with a mass-loading factor $f_{\rm ml}=0.9$, indicating that $90\%$ of the mass falling into the sink is re-injected into the jet. The injection temperature is $T_{\rm jet}=10^8\,{\rm K}$. A key assumption is that the jet axis is always aligned with the instantaneous SMBH spin vector. Any change in the spin direction therefore maps directly onto a change in the jet direction. In order to reduce spurious variations in the injected mass due to grid alignment effects as the AGN jet axis reorients, we replace the classic binary (on/off) selection of jet injection cells---defined by a Boolean mask for a cylindrical injection region---with a smooth transition scheme, in which cells near the radial and axial boundaries of the injection cylinder receive a fractional contribution. A full derivation of the efficiency model and injection procedure is given in P26a.

\subsection{SMBH spin model and evolution} 
\label{sec:spin}

This work uses exclusively the \textit{Hybrid} spin-evolution model introduced and validated in P26a. The dimensionless SMBH spin vector is defined as
\begin{equation}
  \boldsymbol{a} = \frac{c\,\boldsymbol{L}_\bullet}{G M_\bullet^2},
\end{equation}
where $\boldsymbol{L}_\bullet$ and $M_\bullet$ are the black hole angular momentum and mass. The \textit{Hybrid} model combines the direction of the resolved inflow, measured at the sink scale, with a general-relativistic closure for the magnitude of the angular momentum that can physically be deposited onto the hole. In this way, the model retains the three-dimensional variability of the inflow without assuming that the sink-scale specific angular momentum can be transferred directly to the horizon.
The model therefore links the resolved meso-scale inflow to the SMBH through a subgrid horizon closure. It does not resolve the detailed sub-pc circularization flow, Lense--Thirring warping, or viscous alignment; if such unresolved alignment is efficient, the torque direction at the horizon could be smoother than the sink-scale direction used here.

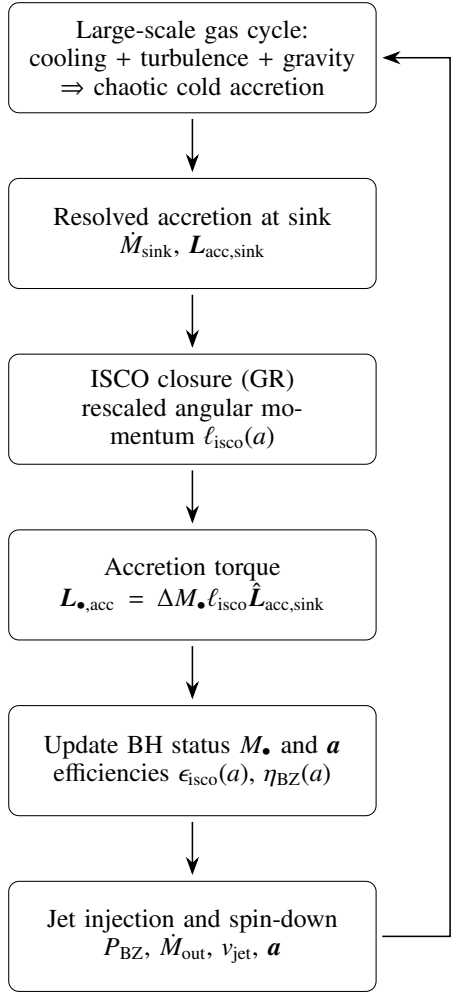
\begin{figure}
\centering
\resizebox{0.65\columnwidth}{!}{%
\begin{tikzpicture}[
  node distance=0.88cm,
  block/.style={draw, rounded corners, align=center, text width=4.7cm, minimum height=1.5cm, minimum width=5cm, inner sep=1pt},
  side/.style={block},
  shared/.style={block},
  arr/.style={-{Stealth[length=2.5mm,width=1.8mm]}, line width=0.6pt, shorten <=2.2pt, shorten >=2.2pt},
  hybrid/.style={block}
]

\node[shared] (s1) {Large-scale gas cycle:\\cooling + turbulence + gravity\\ $\Rightarrow$ chaotic cold accretion};
\node[shared, below=of s1] (s2) {Resolved accretion at sink\\ $\dot{M}_\mathrm{sink}$, $\boldsymbol{L}_{\mathrm{acc,sink}}$};
\node[hybrid, below=of s2] (h2) {ISCO closure (GR)\\ rescaled angular momentum $\ell_\mathrm{isco}(a)$};
\node[hybrid, below=of h2] (h3) {Accretion torque\\ $\boldsymbol{L}_\mathrm{\bullet, {\rm acc}} = \Delta M_{\bullet} \ell_{\text{isco}} \hat{\boldsymbol{L}}_{\text{acc,sink}}$};

\node[shared, below=of h3] (s3) {Update BH status $M_\bullet$ and $\boldsymbol{a}$\\ 
efficiencies $\epsilon_\mathrm{isco}(a)$, $\eta_\mathrm{BZ}(a)$};
\node[shared, below=of s3] (s4) {Jet injection and spin-down\\ $P_\mathrm{BZ}$, $\dot M_\mathrm{out}$, $v_\mathrm{jet}$, $\boldsymbol{a}$};

\draw[arr] (s1.south) -- (s2.north);
\draw[arr] (s2.south) -- (h2.north);
\draw[arr] (h2.south) -- (h3.north);
\draw[arr] (h3.south) -- (s3.north);
\draw[arr] (s3.south) -- (s4.north);
\draw[arr] (s4.east) -- ++(1.0cm,0) |- (s1.east);

\end{tikzpicture}%
}
\caption{Flowchart of the SMBH accretion--spin--feedback coupling used in the simulations, introduced in P26a.}
\label{spin_model}
\end{figure}
At each time step, the model proceeds as follows:

\begin{enumerate}

\item The direction of the angular momentum accreted onto the black hole, $\hat{\boldsymbol{L}}_{\rm \bullet, acc}$, is measured from the resolved flow, summing up the gas angular momentum falling into the sink cells at each time step.

\item The relative orientation between the current spin and the incoming angular momentum, $\hat{\boldsymbol{a}}\cdot\hat{\boldsymbol{L}}_{\rm acc}$, determines whether the ISCO closure is evaluated in the prograde or retrograde branch. If a continuous retrograde accretion episode spins the black hole down to zero, the spin axis flips and the accretion becomes prograde.

\item The mass accreted by the black hole is defined as 
\begin{equation}
\Delta M_{\bullet} = \epsilon_{\text{isco}} \left(1 - f_\mathrm{ml}\right) \dot{M}_{\text{sink}}\Delta t.
\label{eq_mlf}
\end{equation}
The accretion torque is computed using the Kerr specific angular momentum at the ISCO, $\ell_{\rm isco}(a)$, which depends on the spin magnitude and on the ISCO radius \citep[e.g.][]{Bardeen1970, rezzolla2016}. We write
\begin{equation}
\tilde{\ell}_{\rm isco} =
\frac{r_{\rm isco}^2 \mp 2 a \sqrt{r_{\rm isco}} + a^2}
{\sqrt{r_{\rm isco}}(r_{\rm isco}-2) \pm a},
\end{equation}
where the upper sign corresponds to prograde orbits and the lower sign to retrograde orbits (with the corresponding $r_{\rm isco}$). We then set $\ell_{\rm isco}=\tilde{\ell}_{\rm isco}\,GM/c$.
We can then define the angular momentum accreted at each time step by the black hole as
\begin{equation}
\boldsymbol{L}_{\bullet, \text{acc}} = \Delta M_{\bullet} \ell_{\text{isco}} \hat{\boldsymbol{L}}_{\text{acc, sink}}.
\end{equation}
The total SMBH angular momentum is then updated accordingly
\begin{equation}
\boldsymbol{L}_{\bullet, \text{temp}} = \boldsymbol{L}_{\bullet, \text{old}} + \boldsymbol{L}_{\bullet, \text{acc}}.
\end{equation}

\item After applying the subgrid ISCO energy and the jet mass-loading prescriptions, we compute the rotational energy extracted by the jet from the hole through a Blandford--Znajek spin-down torque. Using the horizon angular frequency
\begin{equation}
  \Omega_{\rm H} = \frac{a c^3}{2 G M_\bullet\left(1+\sqrt{1-a^2}\right)},
\end{equation}
we write the corresponding torque magnitude as
\begin{equation}
  \tau_{\rm jet} = \frac{P_{\rm BZ}}{\Omega_{\rm H}},
\end{equation}
and we assume that it acts along the spin vector but in the opposite direction.

\item The black-hole mass and angular momentum are updated, with 
\begin{equation}
\boldsymbol{L}_{\bullet, \text{new}} = \boldsymbol{L}_{\bullet, \text{temp}} \left( \frac{|\boldsymbol{L}_{\bullet, \text{temp}}| - \Delta L_{\text{jet}}}{|\boldsymbol{L}_{\bullet, \text{temp}}|} \right).
\end{equation}
The new spin vector $\boldsymbol{a}_{\text{new}}$ is then updated using the new mass and angular momentum
\begin{equation}
\boldsymbol{a}_\mathrm{new} = \frac{c\boldsymbol{L}_{\bullet, \text{new}}}{G(M_\bullet+\Delta M_\bullet)^2}.  
\end{equation}
In practice, the jet torque mainly reduces the spin magnitude, while the direction of the spin is set by the angular momentum carried by the accreted gas.
\end{enumerate}

Figure \ref{spin_model} summarizes this accretion--spin--feedback loop. For the present paper, the key point is that the \textit{Hybrid} model allows us to separate scalar feeding from vector torque delivery. A non-zero sink accretion rate does not necessarily imply efficient spin reorientation: if the direction of the accreted angular momentum varies rapidly, successive torque increments cancel and the net change in $\boldsymbol{L}_\bullet$ remains small. The relevant question is therefore not only how much mass reaches the sink, but also whether the associated angular momentum remains coherent long enough to tilt the SMBH spin.

\section{Results}
\label{sec:results}

\begin{figure}[!t]
\includegraphics[width=\columnwidth]{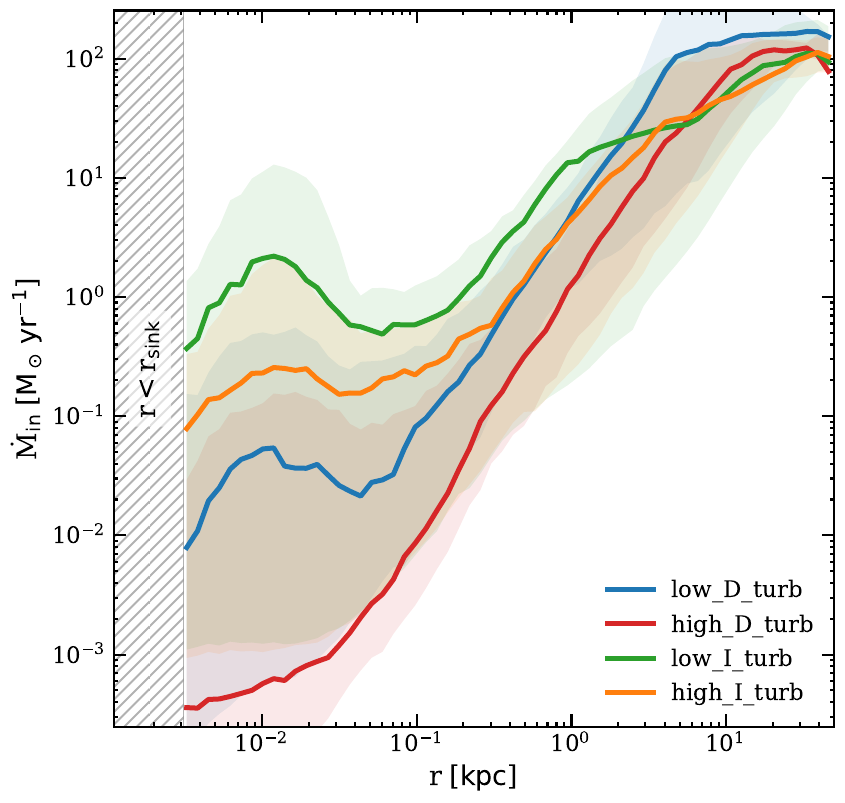}
\caption{Time-averaged radial profile of the total inflow rate (in logarithmic space). The four runs converge to $\dot{M}_{\rm in}\sim10^2\,M_\odot\,{\rm yr^{-1}}$ at large radii, but by $r\sim10\,{\rm pc}$ the IT runs still retain $\sim0.3$--$2\,M_\odot\,{\rm yr^{-1}}$ whereas the DT runs drop to $\sim10^{-3}$--$10^{-2}\,M_\odot\,{\rm yr^{-1}}$. Shaded bands show the $1\sigma$ scatter around the mean.}
\label{mdot_radial_in}
\end{figure}

\begin{figure}[!t]
\includegraphics[width=\columnwidth]{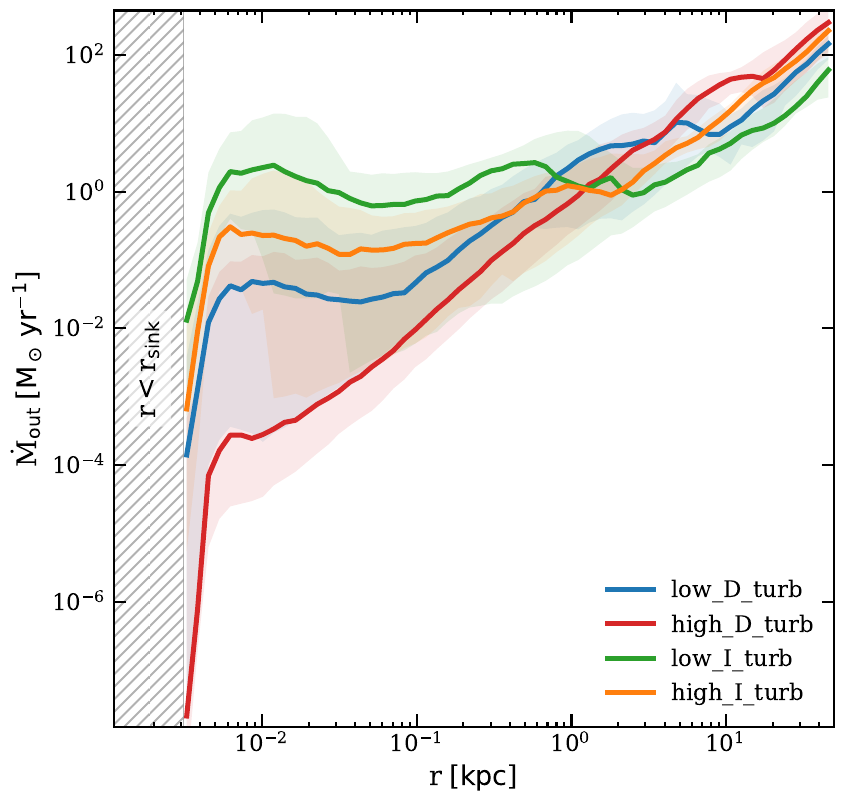}
\caption{Time-averaged radial profile of the total outflow rate (in logarithmic space). The outflow ordering is less monotonic than the inflow ordering because the broader rainy phases of the IT suite also power stronger central feedback events; near $r\sim10$--$100\,{\rm pc}$ the profiles span roughly $10^{-3}$--$1\,M_\odot\,{\rm yr^{-1}}$ across the four runs. Shaded bands show the $1\sigma$ scatter around the mean.}
\label{mdot_radial_out}
\end{figure}

\begin{figure*}[h!]
\includegraphics[width=0.5\textwidth]{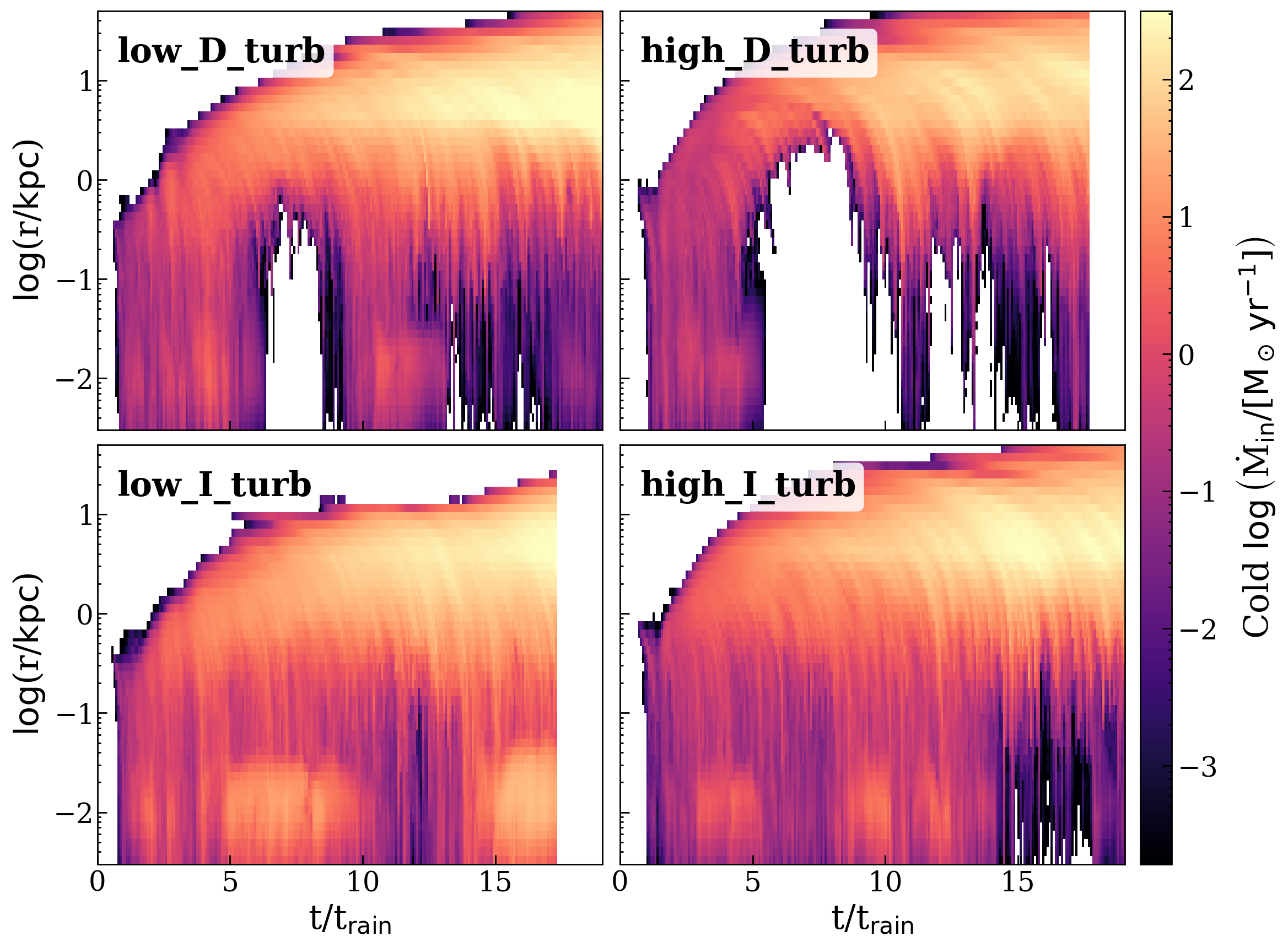}
\includegraphics[width=0.5\textwidth]{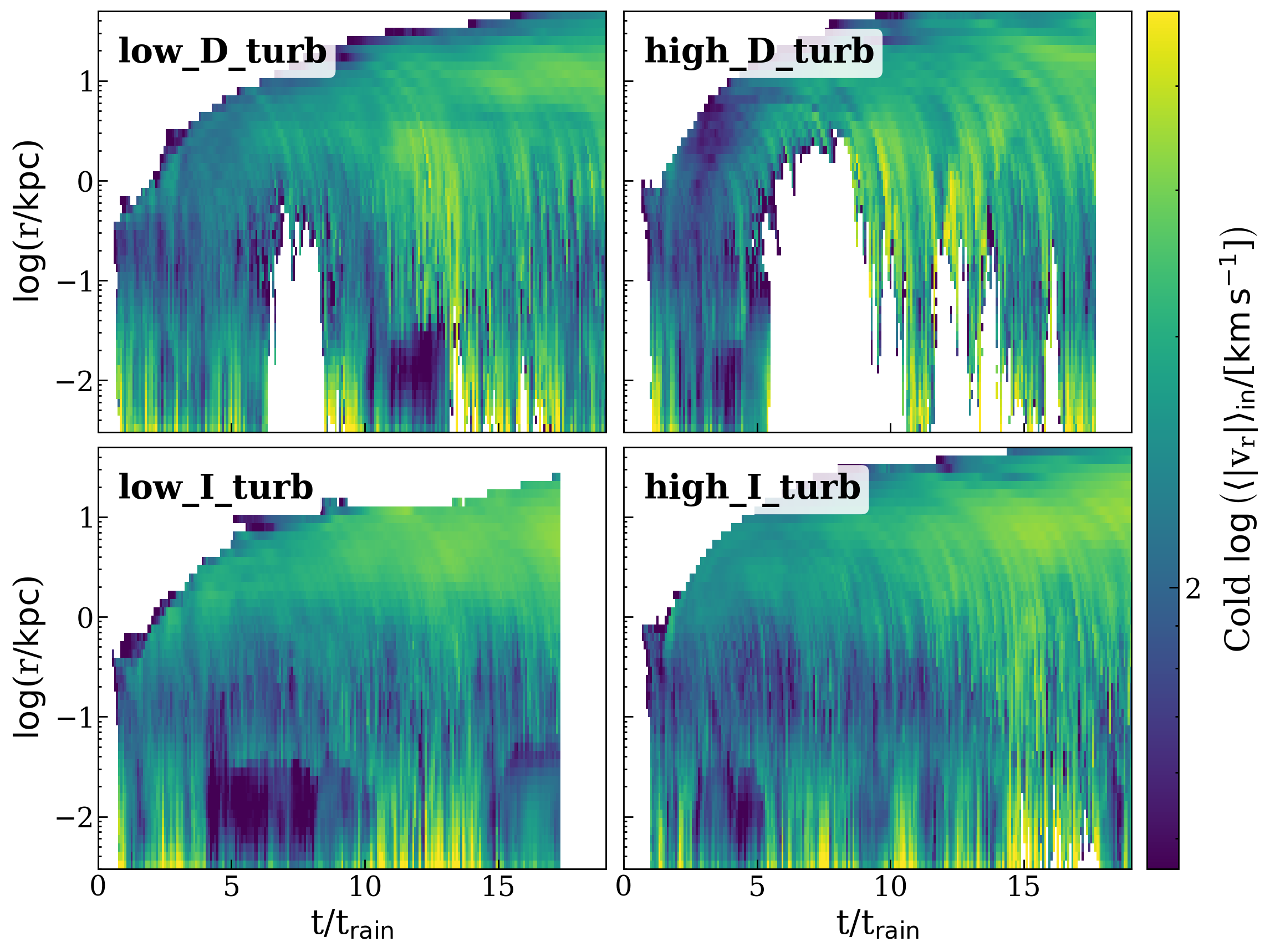}
\caption{Time--radius maps for the four runs of the gross cold inflow mass rate and average inflow radial velocity computed for 50 different radial bins and sampled every 1 Myr. The cold phase provides the clearest view of the different feeding regimes: in the DT suite the inward channel is repeatedly broken at $r\sim0.03$--$2\,{\rm kpc}$ -- what we refer to as \emph{sunny} weather -- while the IT suite maintains a much more continuous cold feeding bridge -- \emph{rainy} weather -- down to the sink.}
\label{inflow_map_cold_main}
\end{figure*}

\begin{figure*}[h!]
\includegraphics[width=0.5\textwidth]{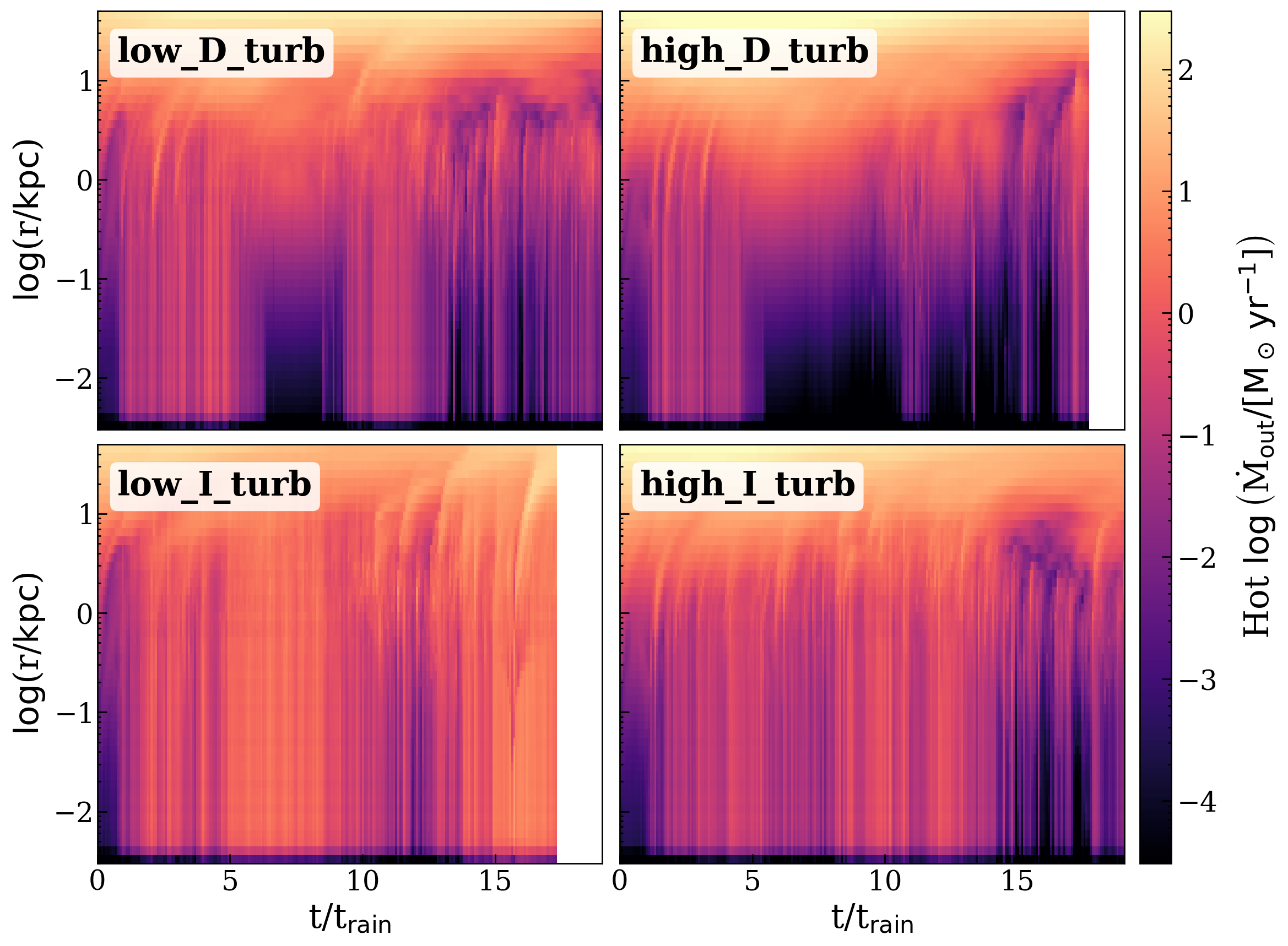}
\includegraphics[width=0.5\textwidth]{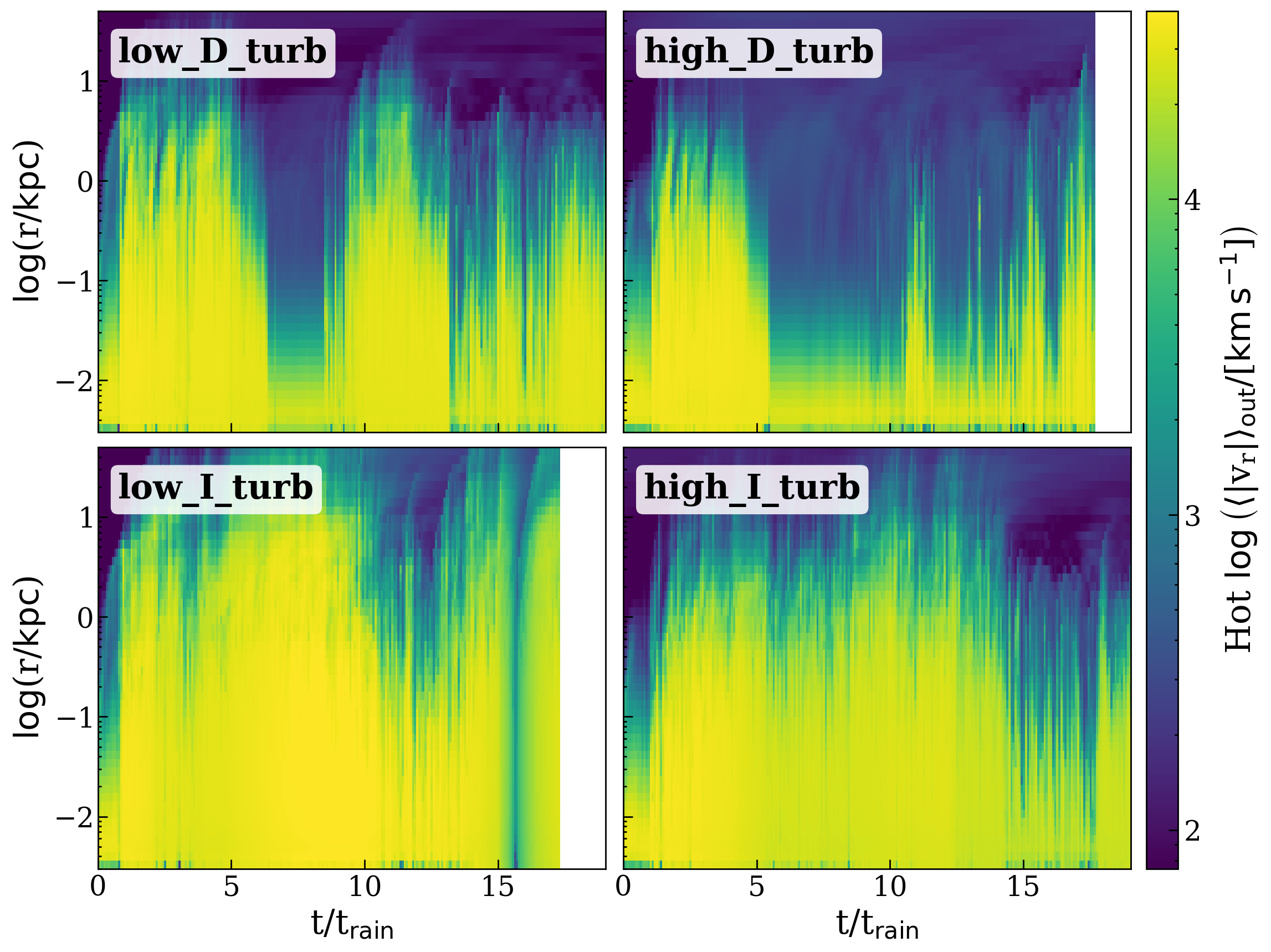}
\caption{Time--radius maps for the four runs of the gross hot outflow mass rate and average outflow radial velocity computed for 50 different radial bins and averaged over 1 Myr. The hot component provides the cleanest view of sunny weather and feedback clearing: all runs launch fast hot outflows, but the IT suite sustains broader post-rain high-velocity episodes, with $|v_{r,{\rm out}}|\gtrsim10^3\,{\rm km\,s^{-1}}$ at the smallest radii.}
\label{outflow_map_hot_main}
\end{figure*}

We now test how ongoing stirring reshapes the nuclear feeding state and how this change is encoded in the SMBH torque statistics, effective jet-axis reorientation, accretion variability, and multiphase halo kinematics. The central question is whether the meso-scale CCA flow preserves radial mass and angular-momentum continuity long enough for successive torque increments to add coherently rather than cancel before reaching the black hole.
Following this sequence, Figures~\ref{mdot_radial_in}--\ref{outflow_map_hot_main} show how the cooling, inflow, and outflow cycles differ across the suite; Figures~\ref{torque} and \ref{prec_rate} connect these weather states to torque coherence and effective jet-axis reorientation; Figure~\ref{psd} shows the associated temporal imprint in the accretion history; and Figures~\ref{cratio}--\ref{kplot_interr} show the corresponding large-scale kinematic differences through key CCA diagnostics.



\subsection{Multiphase mass inflow and outflow}
\label{sec:results_feeding}

To analyze the results, for convenience, we re-normalize time as
\begin{equation}
  \tau \equiv (t-t_\mathrm{on})/t_{\rm rain},
\end{equation}
where $t_{\rm rain} = 16$ Myr is the characteristic precipitation timescale of the first non-linear cooling event in our runs and $t_\mathrm{on} = 50$ Myr is the jet and cooling activation time, so that $\tau$ is set to zero at 50 Myr, and is only used as a normalization to compare different evolutionary stages. We begin with the feeding properties of the four simulations. Figures~\ref{mdot_radial_in} and \ref{mdot_radial_out} show the time-averaged radial profiles of the gross inflow and outflow rates $\dot{M}_\mathrm{in}$ and $\dot{M}_\mathrm{out}$ calculated over all phases as a function of radius radius, while the corresponding Figures \ref{inflow_map_cold_main} and \ref{outflow_map_hot_main} (and in Appendix \ref{app:maps}) resolve the inflow and outflow histories, for the different gas phases, in the time--radius plane. Together, these diagnostics show that the simulations populate distinct multi-scale feeding and weather states.

The clearest trend in this controlled four-run suite is that persistent stirring reduces the radial continuity of the gross inward flux toward the SMBH.
In Figure~\ref{mdot_radial_in}, at large radii, $r\gtrsim10\,{\rm kpc}$, all four runs converge to gross inflow rates of the order $10^2\,M_\odot\,{\rm yr^{-1}}$. The divergence appears inside the central kiloparsec. By $r\simeq1\,{\rm kpc}$ the inflow spans roughly $\sim1$--$15\,M_\odot\,{\rm yr^{-1}}$, and by $r\simeq10\,{\rm pc}$ the spread has widened to roughly 2--3 dex between runs with driven versus interrupted turbulence: \texttt{low\_I\_turb} still retains $\sim1$--$2\,M_\odot\,{\rm yr^{-1}}$, \texttt{high\_I\_turb} retains $\sim0.2$--$0.4\,M_\odot\,{\rm yr^{-1}}$, while \texttt{low\_D\_turb} and especially \texttt{high\_D\_turb} fall to about $10^{-2}$ and $10^{-3}\,M_\odot\,{\rm yr^{-1}}$, respectively. Relative to the common large-radius gross inward flux, this corresponds to only a $\sim2$ dex reduction in the IT suite, but a $\sim4$--5 dex reduction in the DT suite by the time the gas reaches the innermost resolved scales. The relevant effect of persistent stirring is therefore the progressive loss of inward mass-flux continuity and angular-momentum coherence on the way to the sink, which lowers the average delivered accretion rate. In this sense, the dominant bifurcation in the suite is driven (DT) versus interrupted (IT) turbulence, more than low versus high: within each matched pair, switching off the forcing raises the delivered inflow at $r\sim10\,{\rm pc}$ by roughly 2--4 dex.

The outflow profiles are also different. The IT runs show broader high-accretion phases, allowing more gas to reach the hole and therefore also powering stronger central outflow episodes. Near $r\sim10$--$100\,{\rm pc}$ the outflow profiles span approximately $10^{-3}$--$1\,M_\odot\,{\rm yr^{-1}}$, with \texttt{low\_I\_turb} driving a strong centrally concentrated outflow. 
At large radii, the gross inward and outward radial fluxes are similar in magnitude, consistent with a circulation-dominated turbulent halo rather than a simple one-way supply flow. The profiles diverge most strongly only in the central region, \(r\lesssim10\,{\rm pc}\), where the feeding of the sink and the launching of the jet directly affect the flow.
At small radii, around \(\sim10\,{\rm pc}\), the approximate balance \(\dot{M}_{\rm in}\sim\dot{M}_{\rm out}\) suggests a circulation-dominated or partially circularized region with low net radial transport, consistent with the torus-like structures discussed in P26a.

In the next plots -- and in Appendix \ref{app:maps} -- we decompose these gross inward and outward fluxes in time and between the different phases. 
For the time--radius maps we group the five thermodynamic bins into three broader phases: cold gas, combining molecular and cold atomic material with \(T_{\rm cell}<1.6\times10^4\,{\rm K}\); warm gas, with \(1.6\times10^4\,{\rm K}\leq T_{\rm cell}<1.16\times10^6\,{\rm K}\); and hot gas, with \(T_{\rm cell}\geq1.16\times10^6\,{\rm K}\).
For each phase, we then compute the gross inward and outward mass rates $\dot{M}_\mathrm{in}$ and $\dot{M}_\mathrm{out}$, and the average radial velocities $\langle |v_\mathrm{r}|\rangle_\mathrm{in}$ and $\langle |v_\mathrm{r}|\rangle_\mathrm{out}$, indicated with the color bars, for 50 different radial bins, sampled every 1 Myr.

The time--radius maps of the cold inflow phase in Figure \ref{inflow_map_cold_main} clarify the different feeding regimes found in the different runs. 
In \texttt{low\_D\_turb}, the cold, inward channel is repeatedly interrupted at $r\lesssim0.1\,{\rm kpc}$ after $\tau\sim6$, and in \texttt{high\_D\_turb} the nucleus spends much of $\tau\sim7$--17 being fed only through narrow, short-lived inflow threads. Both runs transition from a high-accretion, rainy state to a hot-nucleus, turbulence-dominated sunny phase at $\tau > 6$. By that time, cold gas has already built up at larger scales, which falls towards the centre and helps rebuild the central feeding bridge, connecting the sink to the cold gas now present in the outer shells of the domain (stormy phase). Subsequent feedback events are not able to disrupt the large-scale gas reservoir any more, but can occasionally cut the feeding bridge in the very centre of the box (see $\tau \sim 15$, cloudy phase), reducing the accretion rate again. The interrupted runs (IT) remain variable, but they preserve a much broader and more contiguous inflow bridge from $r\sim1$--$10\,{\rm kpc}$ to $r\lesssim10\,{\rm pc}$ over most of $\tau\sim2$--14, although it starts breaking down at later times as the velocity dispersion continues to rise slowly. The velocity maps tell a similar story. The DT suite shows more discontinuous cold accretion threads with inward speeds of a few $100\,{\rm km\,s^{-1}}$, while the IT suite maintains wider connected regions with typical inward speeds $|v_{r,{\rm in}}|\sim50$--$200\,{\rm km\,s^{-1}}$. The key difference is therefore not a systematically larger inflow speed, but a more persistent radial connection between halo condensation and sink feeding.

The same DT/IT contrast appears in the hot outflow maps of Figure~\ref{outflow_map_hot_main}. Since the jet mass-loading factor is fixed, stronger sink accretion naturally produces stronger jet-driven hot outflows. The IT runs therefore sustain broader and more persistent outflow episodes, whereas the DT runs show stronger intermittency and cycle-to-cycle fragmentation, reflecting the repeated disruption of the cold/warm feeding bridge.
In this sense, the AGN duty cycle in these simulations emerges from how halo stirring regulates the continuity of the multiphase bridge between halo condensation and SMBH feeding. The following sections show that this same continuity also controls whether the delivered angular momentum adds coherently or cancels before it can reorient the SMBH.

\subsection{Torque coherence}
\label{sec:results_torque}

\begin{figure}
\includegraphics[width=0.93\columnwidth]{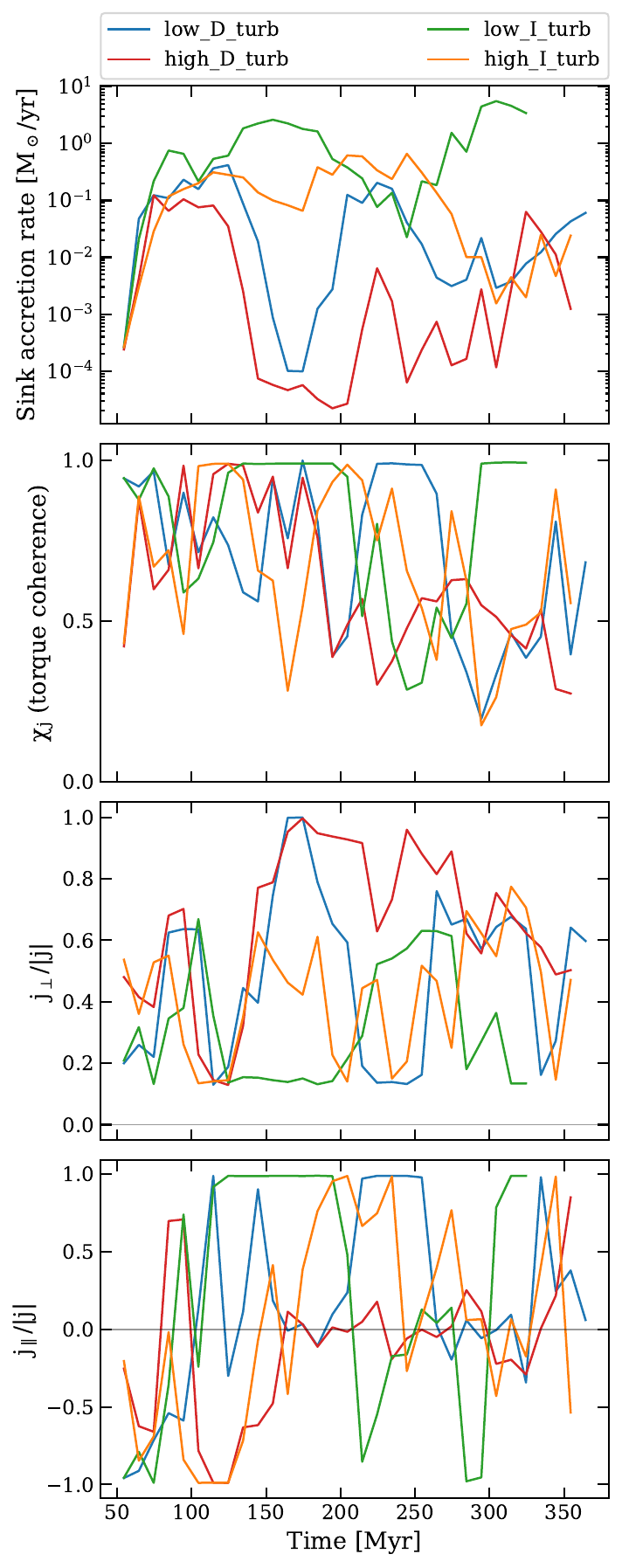}
\vspace{-0.2cm}
\caption{Evolution of the sink accretion rate, torque-coherence parameter, and parallel/perpendicular components of the accreted specific angular momentum, averaged over a 10\,Myr window; shaded bands show the \(1\sigma\) scatter. The IT runs, especially \texttt{low\_I\_turb}, sustain high \(\dot{M}_{\rm sink}\sim0.1\)--\(1\,M_\odot\,{\rm yr^{-1}}\) together with high coherence, often \(\chi_j\gtrsim0.8\), over extended intervals. The DT runs more often decouple coherence from sustained feeding: \texttt{low\_D\_turb} can retain coherent episodes but at lower accretion duty cycle, while \texttt{high\_D\_turb} spends more time in cancellation-dominated states with \(\chi_j\sim0.3\)--0.6.}
\label{torque}
\vspace{-0.2cm}
\end{figure}

To understand how the different feeding states couple to the SMBH, one must go beyond the scalar accretion rate and ask whether the delivered angular momentum adds up coherently or is largely erased by vector cancellation. Figure~\ref{torque} addresses this point. 

The upper panel first shows that the sink accretion histories separate into two broad classes. 
The IT runs sustain $\dot{M}_{\rm sink}\sim0.1$--$1\,M_\odot\,{\rm yr^{-1}}$ for long intervals up to 200-300 Myr, with \texttt{low\_I\_turb} reaching up to $10\,M_\odot\,{\rm yr^{-1}}$. By contrast, \texttt{high\_D\_turb} collapses already after $\sim70-80$ Myr to $\sim10^{-5}$--$10^{-4}\,M_\odot\,{\rm yr^{-1}}$ and thereafter recovers only in short bursts, while \texttt{low\_D\_turb} also drops by roughly three orders of magnitude before re-brightening after a quiescent period of $\sim50\,{\rm Myr}$. This confirms that accretion inflows are more persistent in the IT suite, as expected from the broader rainy phases identified above. 

The decisive sink-scale difference, however, is the torque coherence of the accreting material, shown in the middle panel. 
We define the torque-coherence parameter, which we show averaged over a trailing window $W = 10$ Myr in order to highlight long-term trends more than short-term oscillations:
\begin{equation}
  \chi_j(t;W) \equiv
  \frac{\left|\sum_{t_i\in[t-W,t]} \boldsymbol{L}_{\bullet, \text{acc}}(t_i)\right|}
       {\sum_{t_i\in[t-W,t]} \left|\boldsymbol{L}_{\bullet, \text{acc}}(t_i)\right|}.
\end{equation}
Values \(\chi_j\simeq1\) indicate that the delivered torque keeps an almost fixed direction over the window \(W\), while smaller values indicate stronger vector cancellation from rapidly varying inflow directions. During the rainy intervals of the IT runs, especially \texttt{low\_I\_turb}, the coherence remains high, frequently \(\chi_j\gtrsim0.8\)--1 over several-Myr windows. The DT suite behaves differently: \texttt{low\_D\_turb} can still retain coherent episodes, but at a lower accretion duty cycle, while \texttt{high\_D\_turb} spends more time in cancellation-dominated states with \(\chi_j\sim0.3\)--0.6. Persistent stirring therefore does not simply randomize the torque instantaneously; it shortens the intervals over which a coherent angular-momentum direction remains radially connected to significant sink feeding. In physical terms, continuous turbulence disrupts the longevity of coherent inflow channels or transient partially circularized structures whose angular momentum could otherwise remain aligned or anti-aligned with the SMBH spin and deliver cumulative prograde or retrograde torque.

To determine whether coherent accretion episodes are prograde or retrograde relative to the instantaneous SMBH spin, we decompose the specific angular momentum of the accreted gas into a perpendicular and parallel component, defining
\begin{equation}
  \boldsymbol{j}_{\rm acc}(t) \equiv \frac{\boldsymbol{L}_{\bullet, \text{acc}}(t)}{\Delta M_\bullet(t)}.
\end{equation}
If $\hat{\boldsymbol{a}}(t) \equiv \boldsymbol{a}(t)/|\boldsymbol{a}(t)|$ is the spin unit vector, we define the signed component parallel to the spin,
\begin{equation}
  j_{\parallel}(t) \equiv \boldsymbol{j}_{\rm acc}(t)\cdot \hat{\boldsymbol{a}}(t),
\end{equation}
and the perpendicular amplitude,
\begin{equation}
  j_{\perp}(t) \equiv \left(|\boldsymbol{j}_{\rm acc}(t)|^2 - j_{\parallel}^2(t)\right)^{1/2}.
\end{equation}
Here $j_{\parallel}>0$ corresponds to prograde delivery, $j_{\parallel}<0$ to retrograde delivery, and large $j_{\perp}$ identifies strongly misaligned torques that can tilt the SMBH spin axis efficiently.
We show the time evolution of both components in the bottom panel of the plot. 
Strongly misaligned episodes are common in all runs: $j_{\perp}/|j|$ (solid lines), always positive by definition, frequently reaches $\sim0.5$--1. At the same time, the sign of $j_{\parallel}/|j|$ flips repeatedly, especially in the more turbulent DT suite, each flip corresponding to a transition from a prograde to a retrograde accretion episode. That repeated sign reversal is precisely how a non-zero sink accretion rate can coexist with a weak net torque on the black hole. This picture is supported by the results shown in Table \ref{table_torque_comp}, which confirms that low-turbulence runs indeed maintain a higher degree of coherence of the delivered torque.

A crucial implication is that misalignment by itself is not enough to drive sustained reorientation. Large $j_{\perp}$ is present in every run, but only when it is embedded in a feeding episode with high $\dot M_{\rm sink}$ and persistent sign memory does it produce a cumulative tilt. The relevant sink-scale control parameter is therefore not the amplitude of instantaneous misalignment alone, but the combination of misalignment, accretion rate, and coherence. 
A useful physical proxy for the subsequent reorientation is therefore the coherent perpendicular torque per unit SMBH angular momentum, rather than \(\chi_j\) alone. In practice, rapid spin-axis changes require the simultaneous presence of high \(\dot{M}_{\rm sink}\), large \(j_\perp\), and a coherence time long enough for successive torque increments to add constructively.

\begin{table*}
\centering
\caption{Statistical parameters for all runs: average sink accretion rate, maximum inclination angle, median torque coherence, coherence duty cycle, retrograde duty cycle, mean Eddington ratio, and Eddington-ratio duty cycle.}\label{table_torque_comp}
\begin{tabular}{lrrrrrrr}
\toprule
Run & $\langle \dot M_{\rm sink}\rangle$ & $\theta_{\rm max}$ & median $\chi_j$ & $f(\chi_j>0.7)$ & $f(j_\parallel<0)$ & $\langle \lambda_\mathrm{Edd} \rangle$ & $f(\lambda_\mathrm{Edd} > 0.01)$ \\
\midrule
\texttt{low\_D\_turb} & 0.0669 & 2.13 & 0.832 & 0.577 & 0.444 & 0.0107 & 0.284 \\
\texttt{high\_D\_turb} & 0.0177 & 1.12 & 0.587 & 0.362 & 0.599 & 0.0028 & 0.112 \\
\texttt{low\_I\_turb} & 1.78 & 178 & 0.981 & 0.76 & 0.433 & 0.267 & 0.806 \\
\texttt{high\_I\_turb} & 0.162 & 4.59 & 0.679 & 0.485 & 0.48 & 0.0257 & 0.505 \\
\bottomrule
\end{tabular}
\end{table*}

\begin{figure}
\includegraphics[width=\columnwidth]{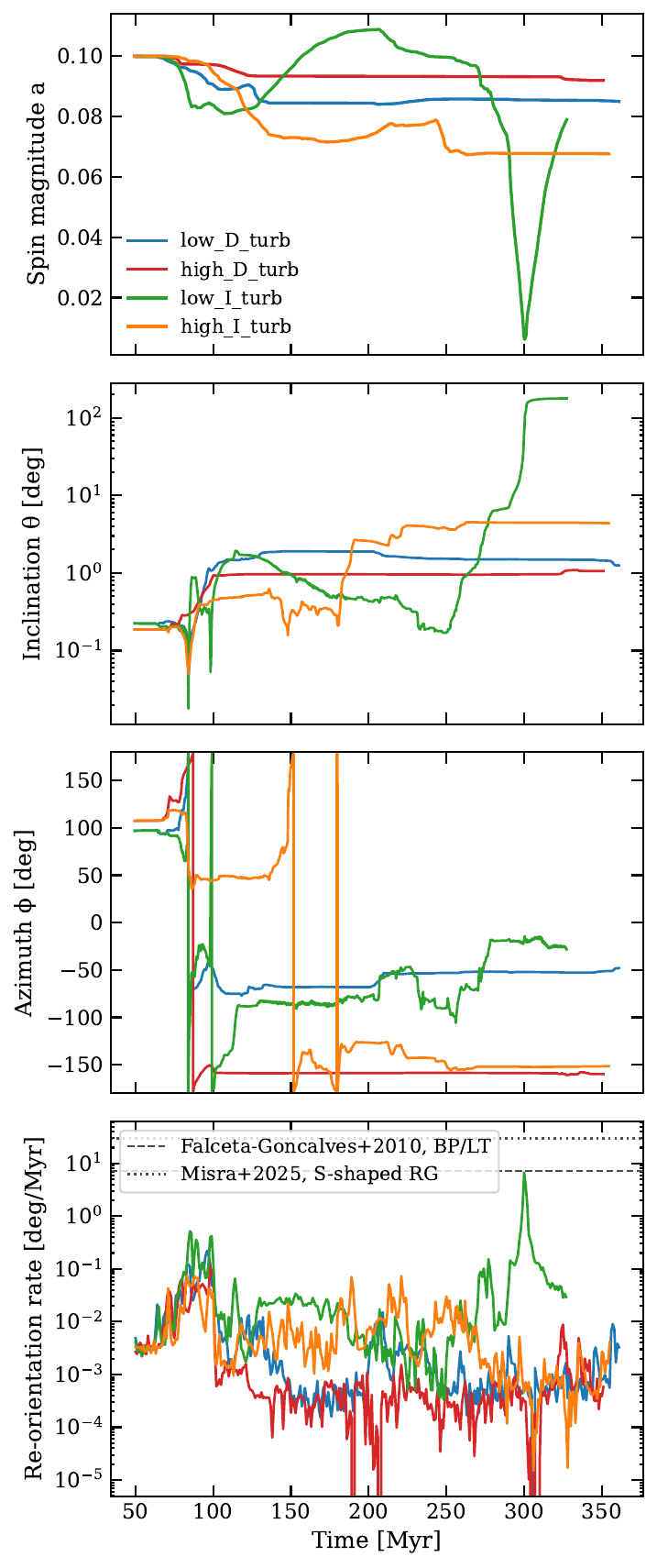}
\vspace{-0.5cm}
\caption{Time evolution of the spin magnitude, inclination with respect to the \(z\)-axis, azimuth, and effective jet-axis reorientation rate. Apart from the near-null reset in \texttt{low\_I\_turb}, all runs remain in the low-spin regime, \(a\sim0.07\)--0.11. After the activation transient, the DT runs settle to slower reorientation rates, \(\sim10^{-4}\)--\(10^{-3}\,{\rm deg\,Myr^{-1}}\), while the IT runs typically remain at \(\sim10^{-2}\)--\(10^{-1}\,{\rm deg\,Myr^{-1}}\), with brief peaks of a few \({\rm deg\,Myr^{-1}}\) during coherent retrograde episodes. The large inclination excursion in \texttt{low\_I\_turb} reflects a spin reset, not smooth periodic precession. Horizontal lines show observational reference ranges from cool-core cavities \protect\citep{falcetagoncalves2010} and S-shaped radio morphologies \protect\citep{misra2025}, although we should be careful with comparisons with observations, since it is not yet clear what the driving physical mechanism of the jet precession is in these cases.}
\label{prec_rate}
\end{figure}

\subsection{Jet-axis re-orientation}
\label{sec:results_spin}

The consequence of the different torque histories is shown in Figure~\ref{prec_rate}, which shows how the feeding state of the nucleus is converted into the dynamical state of the SMBH and, through the jet-alignment prescription, into the outflow orientation history. 
To this end, we define an effective jet-axis reorientation rate as the angular change of the spin direction between successive outputs:
\begin{equation}
  \dot{\theta}_i \equiv \frac{\cos^{-1}\!\left(\hat{\boldsymbol{a}}_i\cdot\hat{\boldsymbol{a}}_{i+1}\right)}{\Delta t_i}.
\end{equation}
We report this quantity in \({\rm deg\,Myr^{-1}}\) and plot it averaged over 1-Myr bins. 
Because the spin evolution in CCA is irregular rather than periodic, \(\dot{\theta}\) should not be interpreted as a formal precession frequency.
It is instead an effective orientational response of the adopted spin--jet closure to weather-regulated torque delivery.

As a general trend, lower sink accretion rates and lower torque coherence produce slower spin evolution. The DT runs show the weakest spin-magnitude response: \texttt{low\_D\_turb} declines from \(a\simeq0.10\) to \(\simeq0.085\) by \(\sim120\,{\rm Myr}\) and then remains nearly flat, while \texttt{high\_D\_turb} stays close to \(a\simeq0.093\). The IT runs evolve more strongly: \texttt{high\_I\_turb} declines to \(a\simeq0.071\), partially recovers to \(\simeq0.078\), and then settles near \(a\simeq0.067\). The most extreme case is \texttt{low\_I\_turb}: after an early decline and subsequent spin-up to \(a\simeq0.11\), a coherent retrograde episode around \(300\,{\rm Myr}\) drives the spin toward a near-null state. The dominant split is therefore DT versus IT, with interrupted turbulence allowing coherent feeding episodes to couple more efficiently to the SMBH spin.

The inclination with respect to the \(z\)-axis is shown in the middle panel. The DT runs remain confined to \(\sim1\)--\(2^\circ\), while \texttt{high\_I\_turb} reaches a few degrees. \texttt{low\_I\_turb} instead undergoes a spin reset: once the spin magnitude is nearly erased, subsequent accretion rebuilds the spin along a new direction, so the formal inclination jumps to a flip-like value. This should not be interpreted as smooth large-amplitude precession. The azimuthal angle can also show large excursions, in some cases exceeding \(\sim100^\circ\), but when the inclination is small such changes mostly reflect motion around the polar axis and do not necessarily imply rapid physical jet-axis reorientation. In this low-spin regime, the strongest orientation changes occur when coherent feeding first reduces the spin angular-momentum reservoir and then rebuilds it along a different direction.

The bottom panel shows the evolution of the effective jet-axis reorientation rate. All runs exhibit an early activation transient once cooling and feedback are turned on. After that transient, the DT runs mostly settle to $\dot{\theta}\sim10^{-4}$--$10^{-3}\,{\rm deg\,Myr^{-1}}$, with only occasional excursions toward $\sim10^{-2}$--$10^{-1}\,{\rm deg\,Myr^{-1}}$, whereas the IT runs remain typically at $\sim10^{-2}$--$10^{-1}\,{\rm deg\,Myr^{-1}}$ and, during the strongest coherent retrograde episodes, can briefly spike to a few ${\rm deg\,Myr^{-1}}$. 
In this sense, the axis-reorientation history is an orientational signature of the feeding state identified earlier: connected, high-feeding, coherent phases drive rapid steering, while fragmented or cancellation-dominated phases drive slow drift.

In a smooth disc, spin evolution would be controlled mainly by the long-term sign and magnitude of the accreted angular momentum. In CCA, instead, the SMBH is fed by a sequence of clouds, filaments, collisions, and fallback streams whose angular momenta can either add coherently or cancel before reaching the horizon closure. The spin therefore behaves as an integrator of the connected feeding bridge. Extended intervals with high \(\dot{M}_\mathrm{sink}\), large \(j_\perp\), and high \(\chi_j\) can steer the jet axis efficiently. On the other hand, cloudy or sunny intervals may still contain multiphase gas but deliver only fragmented, rapidly changing torques. This interpretation should be understood in the context of the low-spin regime explored here, where the SMBH angular-momentum reservoir is modest and therefore easier to reorient; higher-spin SMBHs would require larger or longer-lived coherent torques to achieve the same angular displacement, as already shown in P26a. 

\begin{figure}
\includegraphics[width=\columnwidth]{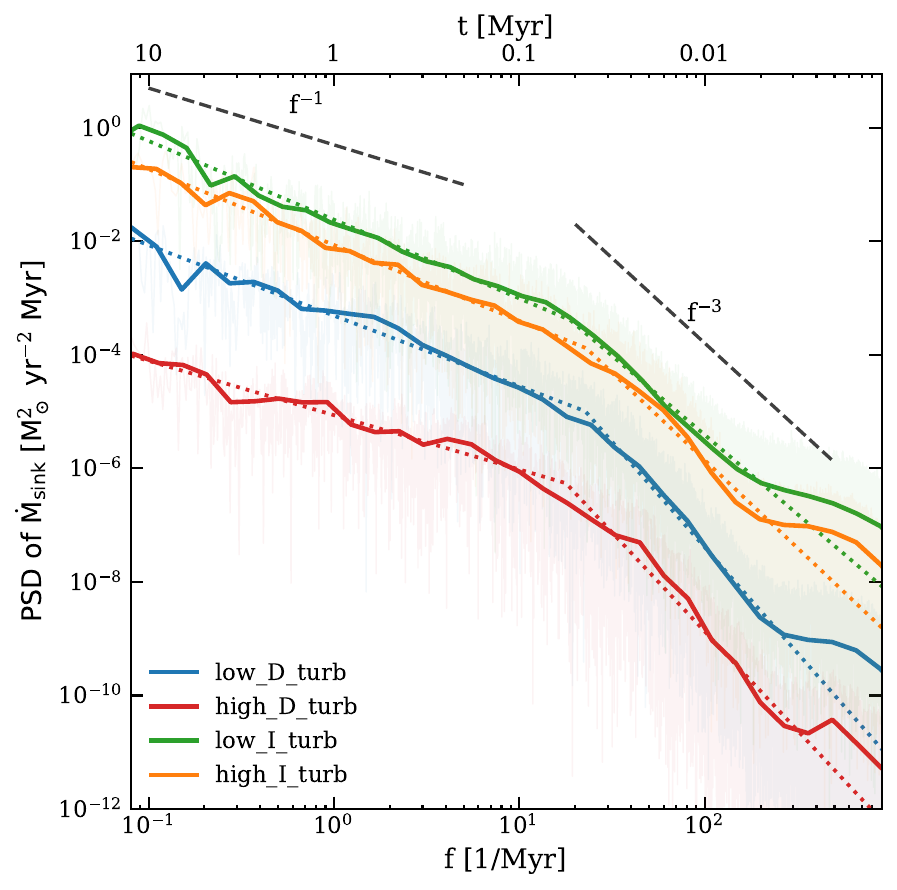}
\caption{Power spectral density (PSD) of the sink accretion rate for the four runs. At $f\sim0.1$--$1\,{\rm Myr^{-1}}$, the IT runs lie roughly 1--2 dex above the DT runs, with \texttt{low\_I\_turb} carrying the largest low-frequency power and \texttt{high\_D\_turb} the smallest. 
The fitted breaks cluster near \(f_{\rm b}\sim20\,{\rm Myr^{-1}}\), corresponding to \(\sim0.05\,{\rm Myr}\), and therefore most likely trace a common inner accretion-response. The DT/IT contrast is instead encoded in the PSD normalization and slopes: IT runs retain enhanced low-frequency power from longer connected rainy episodes, while DT runs show steeper high-frequency damping, consistent with stronger fragmentation and angular-momentum cancellation before the inflow reaches the sink. Dotted lines show the double power law fit (see Table \ref{table_psd}), while the black dashed lines show reference slopes $\propto -1$ and $\propto -3$.}
\label{psd}
\end{figure}
\subsection{Time variability via power spectra}
\label{sec:results_psd}

\begin{table}
\caption{PSD fitting results assuming double power-law shapes. The fit is performed within the frequency range $f = 0.08$--$300\,\mathrm{Myr^{-1}}$. \label{table_psd}}
\centering
\begin{tabular}{lccc}
\toprule
Run & $\alpha_\mathrm{1}$ & $\alpha_\mathrm{2}$ & $\log(f_\mathrm{b})$\\
\midrule
\texttt{low\_D\_turb}  & $1.24 \pm 0.05$ & $3.72 \pm 0.16$ & $1.35 \pm 0.05$\\
\texttt{high\_D\_turb} & $0.96 \pm 0.06$ & $3.49 \pm 0.15$ & $1.25 \pm 0.05$\\
\texttt{low\_I\_turb}  & $1.39 \pm 0.05$ & $2.79 \pm 0.12$ & $1.27 \pm 0.07$\\
\texttt{high\_I\_turb} & $1.34 \pm 0.04$ & $3.09 \pm 0.13$ & $1.36 \pm 0.06$\\
\bottomrule
\end{tabular}
\end{table}

In Figure~\ref{psd} we computed the power spectral density (PSD) of the sink accretion rate. The considered period is restricted to after the cooling and jet onset. The irregularly sampled accretion-rate series is linearly interpolated onto a uniform grid with cadence $\Delta t = 10^{-4}\ {\rm Myr}$. The one-sided PSD was computed from the real FFT using the normalization
\[
P(f_k) = \frac{\Delta t}{N\,U}\left|\dot{M}_{\mathrm{sink}, k}\right|^2,
\]
where \(N\) is the number of uniformly sampled points and $U = \langle w^2\rangle$ is the mean-square value of the (Hann) window function. The PSD was then logarithmically binned in frequency before fitting it with a broken power law,
\[
P(f) =
A
\begin{cases}
(f/f_{\rm b})^{-\alpha_1}, & f < f_{\rm b},\\
(f/f_{\rm b})^{-\alpha_2}, & f \ge f_{\rm b},
\end{cases} 
\] 
by performing a least-squares fit in \(\log_{10} P\) space over the frequency range $0.08 \le f \le 300\ {\rm Myr}^{-1}$. The uncertainties reported are the uncertainties of the formal \(1\sigma\) parameter estimated from the square roots of the diagonal elements of the covariance matrix returned by the nonlinear least-squares optimizer. The fitting parameters obtained are shown in Table \ref{table_psd}.
Figure~\ref{psd} shows that all runs retain flicker-noise accretion variability. Even when the external turbulent driving is interrupted, the system does not settle into a steady or periodic state: radiative precipitation, cloud--cloud interactions, filament fallback, angular-momentum cancellation, and jet feedback continue to generate chaotic feeding over a wide range of timescales. This is the expected temporal behaviour of CCA, where the SMBH is fed by a rain of multiphase condensates rather than by a smooth, laminar inflow or a long-lived coherent disc. 

The main DT/IT difference is the normalization and temporal organization of this stochasticity. Across most of the fitted frequency range, the PSDs follow the ordering
\(\texttt{low\_I\_turb} > \texttt{high\_I\_turb} > \texttt{low\_D\_turb} > \texttt{high\_D\_turb}\), showing that interrupted-turbulence runs retain substantially more accretion power than continuously driven runs. In the CCA picture, this reflects how efficiently large-scale precipitation is transmitted to the sink. When cold/warm gas remains radially connected to the nucleus, chaotic cloud interactions are not erased, but are organized into longer, higher-amplitude rainy feeding episodes. When stirring persists, the same multiphase condensation cascade is more strongly mixed and fragmented; clouds and filaments collide, shear, and cancel angular momentum before reaching the sink, so the variability is filtered into lower-amplitude, more rapidly decorrelating bursts.
At low frequencies, the spectra are broadly consistent with \(P(f)\propto f^{-1}\), similar to the flicker-/pink-noise behaviour expected for chaotic, self-regulated CCA variability \citep[cf.][]{gaspari2017}. The fitted low-frequency slopes span \(\alpha_1\simeq1.0\)--1.4, while the largest physical separation among the runs is the excess low-frequency power in the IT suite, especially in \texttt{low\_I\_turb}. This low-frequency enhancement is the temporal imprint of connected rain: the flow remains chaotic, but rain and infall remain coupled to the nucleus for longer intervals, allowing multiphase feeding events to add coherently in time.

At high frequencies, the DT runs show steeper damping, with \(\alpha_2\simeq3.5\)--3.7 compared with \(\alpha_2\simeq2.8\)--3.1 in the IT suite. This indicates that persistent stirring does not simply inject extra small-scale flickering into the SMBH accretion rate. Instead, by disrupting radial connectivity and enhancing angular-momentum cancellation, it suppresses the transmission of cloud-scale fluctuations to the BH. In this sense, the DT suite behaves like a more efficient stochastic filter between halo precipitation and nuclear feeding, whereas the IT suite preserves a more direct precipitation--accretion connection.

The fitted breaks cluster near \(f_{\rm b}\sim20\,{\rm Myr^{-1}}\), corresponding to \(\sim0.05\,{\rm Myr}\), and should therefore not be interpreted as the timescale of the multi-Myr weather cycle. The robust CCA/weather imprint is instead the combination of enhanced low-frequency power in the IT runs and stronger high-frequency damping in the DT runs. Broad rainy intervals produce long-memory accretion fluctuations, while cloudy or sunny, fragmentation-dominated intervals erase coherence before the gas reaches the SMBH. At \(f\gtrsim300\,{\rm Myr^{-1}}\), the spectra approach a numerical white-noise floor associated with the finite timestep or output cadence. 

Compared with the companion {\sc BlackHoleWeather} papers, the present runs preserve the same low-frequency CCA variability, with slopes close to \(P(f)\propto f^{-1}\), but modify how this variability is transmitted to the SMBH. Relative to B26b, the IT runs show enhanced low-frequency power, while the DT runs show systematically lower PSD normalization due to the persistent stirring of the multiphase flow. Relative to C26b, the DT runs also show stronger high-frequency damping. Thus, the main difference is not the presence of CCA-like stochasticity, but the efficiency with which precipitation-driven fluctuations remain connected to nuclear feeding.

\subsection{Kinematical CCA diagnostics}
\label{sec:results_kinematics}

We finally examine how the different feeding states are associated with halo-scale kinematic differences. These diagnostics provide a large-scale view of the same weather-regulated cycle identified from the sink accretion and torque histories, connecting the nuclear response to the thermodynamic and kinematic state of the multiphase halo. We have already seen in Figure~\ref{sigmav} that the global velocity dispersion evolves differently throughout the suite. 
\texttt{high\_D\_turb} rises rapidly to \(\sim160\,{\rm km\,s^{-1}}\) by \(\sim50\,{\rm Myr}\) and reaches \(\sim300\)--\(315\,{\rm km\,s^{-1}}\) by \(\sim120\,{\rm Myr}\), remaining there thereafter. \texttt{low\_D\_turb} climbs more gradually but still reaches \(\sim200\)--\(210\,{\rm km\,s^{-1}}\) by the end of the run. The IT runs saturate at lower values and later converge toward \(\sim150\)--\(160\,{\rm km\,s^{-1}}\). 

We now use more in-depth kinematic information to probe three key CCA diagnostics \citep[first proposed in][]{gaspari2018}: the condensation ratio \(C\equiv t_{\rm cool}/t_{\rm eddy}\), the turbulent Taylor number \(\mathrm{Ta_t}\equiv v_{\rm rot}/\sigma_v\), and the projected kinematic k-plot.

\begin{figure}
\includegraphics[width=0.94\columnwidth]{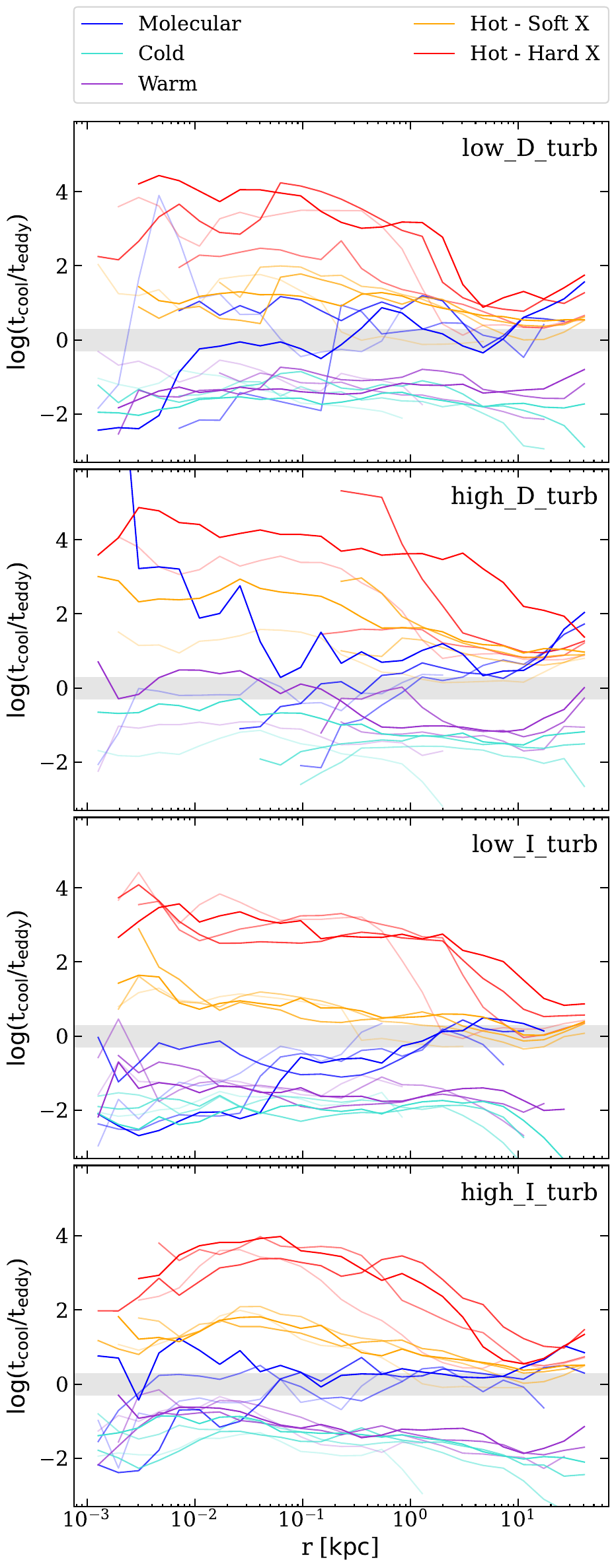}
\vspace{-0.3cm}
\caption{Radial profiles of the condensation ratio \(C\equiv t_{\rm cool}/t_{\rm eddy,hot}\) at selected epochs \((\tau=2,8,14,20;\) increasing transparency with time). The grey band marks the canonical CCA rain range \(C\sim0.5\)--2 \citep{gaspari2018}. The hard-X gas mostly traces the hot turbulent reservoir at \(\log_{10}C>0\), while the soft-X phase approaches the condensation band where warm/cold gas forms. Cooler phases lie at \(\log_{10}C<0\) because they are already nonlinear products of the cooling cascade, and are normalized here to the hot-phase eddy time.}
\vspace{-0.3cm}
\label{cratio}
\end{figure}
\begin{figure}
\includegraphics[width=0.94\columnwidth]{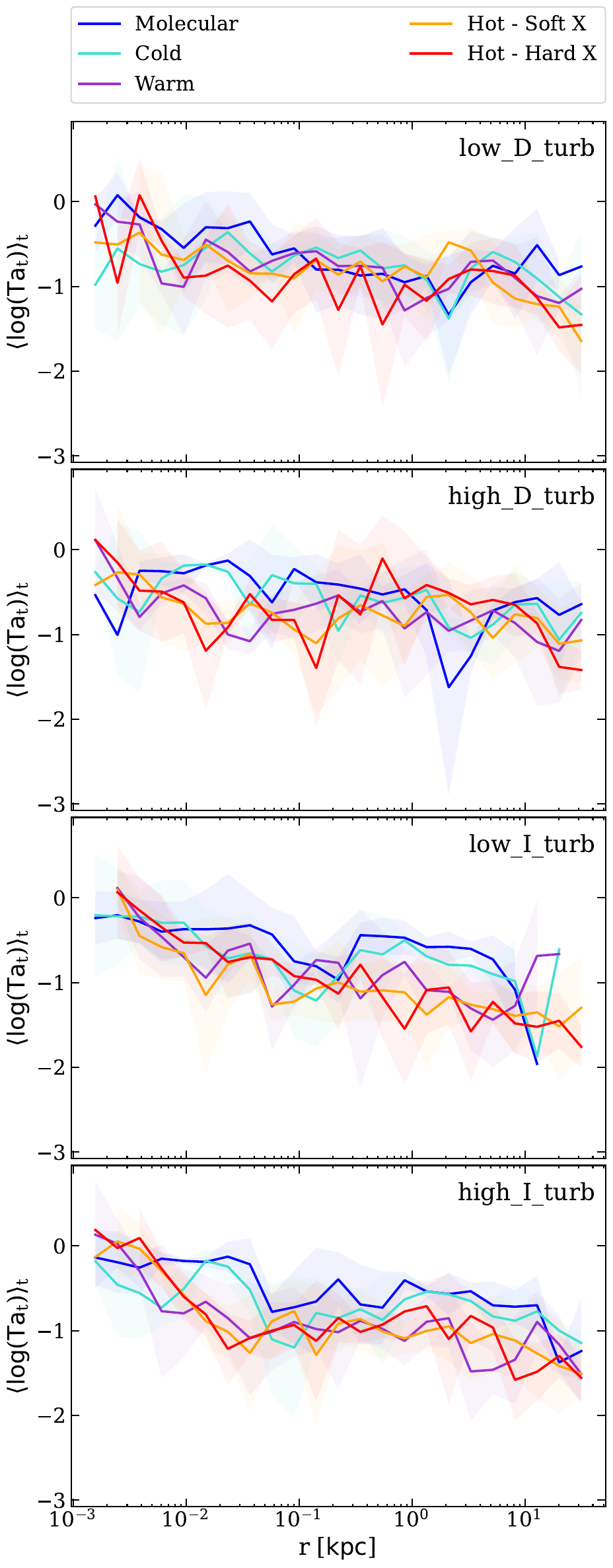}
\caption{Radial profiles of the time-averaged turbulent Taylor number \(\mathrm{Ta_t}(r)\equiv v_{\rm rot}(r)/\sigma_v(r)\) for the four runs, with the shaded areas showing the 1-$\sigma$ deviation from the average. In fact, contrary to the $\mathcal{C}$-ratio, the $\mathrm{Ta_t}$ does not show significant time evolution. Most values lie at $\log_{10}(\mathrm{Ta_t})\lesssim0$, i.e. $\mathrm{Ta_t}\lesssim1$, confirming that the flow remains largely dispersion-dominated (quasi-isotropic CCA rain; \citealt{gaspari2015}) rather than settling into a rotationally supported disc.}
\vspace{-0.3cm}
\label{tat}
\end{figure}

\subsubsection{Condensation ratio}
\label{sec:results_cratio}

Figure~\ref{cratio} shows the radial structure of the condensation ratio, \(C\equiv t_{\rm cool}/t_{\rm eddy}\), separated by gas phase and epoch. This diagnostic measures whether radiative cooling can compete with turbulent mixing and circulation, and has been developed and applied in the CCA framework to connect \(C\sim1\) with the onset and radial extent of multiphase condensation in hot halos \citep{gaspari2018,Maccagni2021,Olivares2022,Wang2023}. Here we use the hot-phase eddy time, \(t_{\rm eddy,hot}\), because the condensation cascade is initiated out of the hot atmosphere before producing the warm, cold, and molecular phases. The most physically informative regime is therefore the hot gas, especially the soft-X component.
Here the cooling time can be approximated as 
\begin{equation}
t_{\rm cool} \simeq \frac{3 k_B T}{n_e\,\Lambda},
\end{equation}
with $T$ being the gas temperature, $n_e$ the electron density, and $\Lambda(T, Z)$ the cooling function. The eddy time quantifies how quickly subsonic turbulence stirs the gas on a given scale
\begin{equation}
t_{\rm eddy} = 2\pi \,\frac{r^{2/3}\,L^{1/3}}{\sigma_{v,L}},
\end{equation}
where $L$ is the turbulence injection scale and $\sigma_{v,L}$ is the velocity dispersion measured at that scale. This expression assumes a Kolmogorov cascade for subsonic turbulence, with $\sigma_v(l)\propto l^{1/3}$ \citep{kolmogorov1941,gasparichurazov2013,fournier2025}. For these plots, we further divide the gas phases into molecular ($T_\mathrm{cell}<2\times10^{2}\ \mathrm{K}$), atomic ($2\times10^{2}\ \mathrm{K}\leq T_\mathrm{cell}<1.6\times10^{4}\ \mathrm{K}$), warm ($1.6\times10^{4}\ \mathrm{K}\leq T_\mathrm{cell}<1.16\times10^{6}\ \mathrm{K}$), soft X-ray ($1.16\times10^{6}\ \mathrm{K}\leq T_\mathrm{cell}<5.8\times10^{6}\ \mathrm{K}$) and hard X-ray gas ($T_\mathrm{cell}>5.8\times10^{6}\ \mathrm{K}$).

Across the suite, the hard-X phase generally lies at high \(\log_{10}C\), especially inside the central \(\sim0.1\)--\(1\,{\rm kpc}\). This gas has a cooling time much longer than the local eddy time and therefore behaves mainly as the ambient turbulent reservoir rather than as the immediate precipitating component. The hard-X curves decrease outward, approaching lower \(C\) at larger radii, but they usually remain above the canonical condensation band. By contrast, the soft-X phase lies much closer to \(C\sim1\), especially over the radii where warm and cold gas subsequently appear. The soft-X gas is thus the clearest tracer of where the hot atmosphere first becomes thermally susceptible to precipitation.

The model-to-model differences add an important layer to this picture. The continuously driven runs, and in particular \texttt{high\_D\_turb}, tend to keep both hard-X and soft-X gas at larger \(C\) in the inner halo. This is expected if persistent stirring shortens \(t_{\rm eddy,hot}\), enhances mixing, and delays the conversion of hot gas into a connected cooling flow. The DT suite is therefore not condensation-free, but its hot atmosphere is more strongly mixing-dominated and the resulting condensate is less efficiently transmitted to the nucleus. The interrupted-turbulence runs, especially \texttt{low\_I\_turb}, show soft-X profiles that remain closer to the \(C\sim1\) band over a broader radial range. This is consistent with the time-radius maps: once external stirring decays, thermally susceptible soft-X gas can feed a more continuous cold/warm precipitation channel toward the sink.

The warm, cold, and molecular phases often lie at \(C\lesssim1\), or equivalently \(\log_{10}C<0\). These phases are, in fact, already downstream products of the cascade: once the gas has cooled and compressed, \(t_{\rm cool}\) becomes short by construction and the ratio drops below unity (see also B26b/C26b). In addition, because all phases are normalized here using \(t_{\rm eddy,hot}\), the cool-phase values should be regarded as a diagnostic of their relation to the hot turbulent background rather than as intrinsic phase-by-phase instability criteria. The important CCA message is that the thermodynamic onset is carried by the soft-X gas, while the colder phases record the nonlinear outcome of that onset.

A further feature visible in Figure~\ref{cratio} is that the molecular component does not always remain deep in the \(C\ll1\) regime. In several runs, especially in the driven suite, the molecular curves approach \(C\sim1\) over parts of the inner halo. This is best interpreted as feedback-processed cold gas rather than as a new molecular condensation threshold. Jet-driven uplift, weak shocks, turbulent mixing, cloud crushing, and fallback can rarefy or partially heat cold clouds and their interfaces, increasing the \(C\)-ratio before the gas either re-cools, mixes into the warm phase, or leaves the molecular bin. This interpretation is consistent with C26b, where AGN feedback with no spin is included. However, the contrast with the driven-turbulence, no-jet simulations of B26b is significant: in those runs, the molecular phase remains at \(C\sim10^{-2}\)--\(10^{-3}\), whereas here it can approach \(\log_{10}C\sim0\). This suggests that AGN-driven uplift, mixing, and partial reheating keep part of the coldest gas dynamically coupled to the hot turbulent atmosphere, rather than allowing it to remain as a purely cooling-dominated nuclear reservoir.

Thus, both DT and IT suites can reach the precipitation regime and form multiphase gas. The difference lies in how the hot-halo thermodynamic susceptibility couples to meso-scale transport: in the IT runs, soft-X gas closer to \(C\sim1\) is more readily converted into connected rain, while in the DT runs persistent stirring keeps the hot phase more mixing-dominated and processes the cold reservoir through uplift, mixing, and partial reheating. The condensation cascade is therefore still active in the driven suite, but it is converted into a fragmented and partially decorrelated multiphase circulation before the gas reaches the SMBH.

\subsubsection{Turbulent Taylor number}
\label{sec:results_tat}

Figure~\ref{tat} shows the turbulent Taylor number \citep{gaspari2015}, \(\mathrm{Ta_t}(r)\equiv v_{\rm rot}(r)/\sigma_v(r)\), which compares ordered rotation with turbulent or dispersion support. This diagnostic was introduced in the CCA context to distinguish the turbulent-rain regime, \(\mathrm{Ta_t}<1\), from rotation-dominated states in which a centrifugal barrier or warm disc can suppress chaotic infall \citep{gaspari2015}. Observationally, the same criterion has been used to interpret rotating early-type galaxies, where systems with \(C\sim1\) but \(\mathrm{Ta_t}>1\) are expected to condense along non-radial orbits and form extended multiphase discs rather than thin CCA filaments \citep[e.g.][]{Juranova2019,Juranova2020}. 
For each radial bin and thermodynamic phase, we compute \(v_{\rm rot}=|\langle v_\phi\rangle_m|\) and compare it with the velocity dispersion used in Figure \ref{cratio}.

Across most radii, phases, and epochs, our simulations remain at \(\mathrm{Ta_t}\lesssim1\), confirming that the flow is dispersion-dominated. This is true for both the continuously driven and interrupted-turbulence suites. The multiphase gas is therefore driven by turbulent condensation, cloud--cloud collisions, uplift, fallback, and chaotic angular-momentum cancellation. In this respect, both the DT and IT suites remain in the CCA regime.

The model-to-model differences instead appear in the degree of scatter and radial coherence of \(\mathrm{Ta_t}\). The DT runs, especially \texttt{high\_D\_turb}, show broader fluctuations and more local excursions around \(\mathrm{Ta_t}\sim1\), consistent with a stirred inner halo with a perturbed multiphase gas. The IT runs show a cleaner decline toward \(\mathrm{Ta_t}<0.1\) outside the innermost region, consistent with a more dispersion-dominated precipitation flow once the external forcing decays. 

The occasional excursions to \(\mathrm{Ta_t}>1\), especially in the IT runs and mostly inside the central \(\sim10\,{\rm pc}\), are consistent with transient, torus-like circularized structures near the sink. They occur in all runs and do not define the global feeding mode. This is important for the spin interpretation: effective jet-axis reorientation depends on whether the meso-scale multiphase bridge remains connected and coherent long enough to deliver cumulative torque to the SMBH. In the IT runs, decaying turbulence allows cold/warm gas to preserve this connection for longer intervals. In the DT runs, continuous stirring keeps the flow more fragmented and cancellation-prone, so the gas remains multiphase but delivers less coherent mass and angular momentum to the black hole.

\begin{figure*}
\centering
\includegraphics[width=0.95\textwidth]{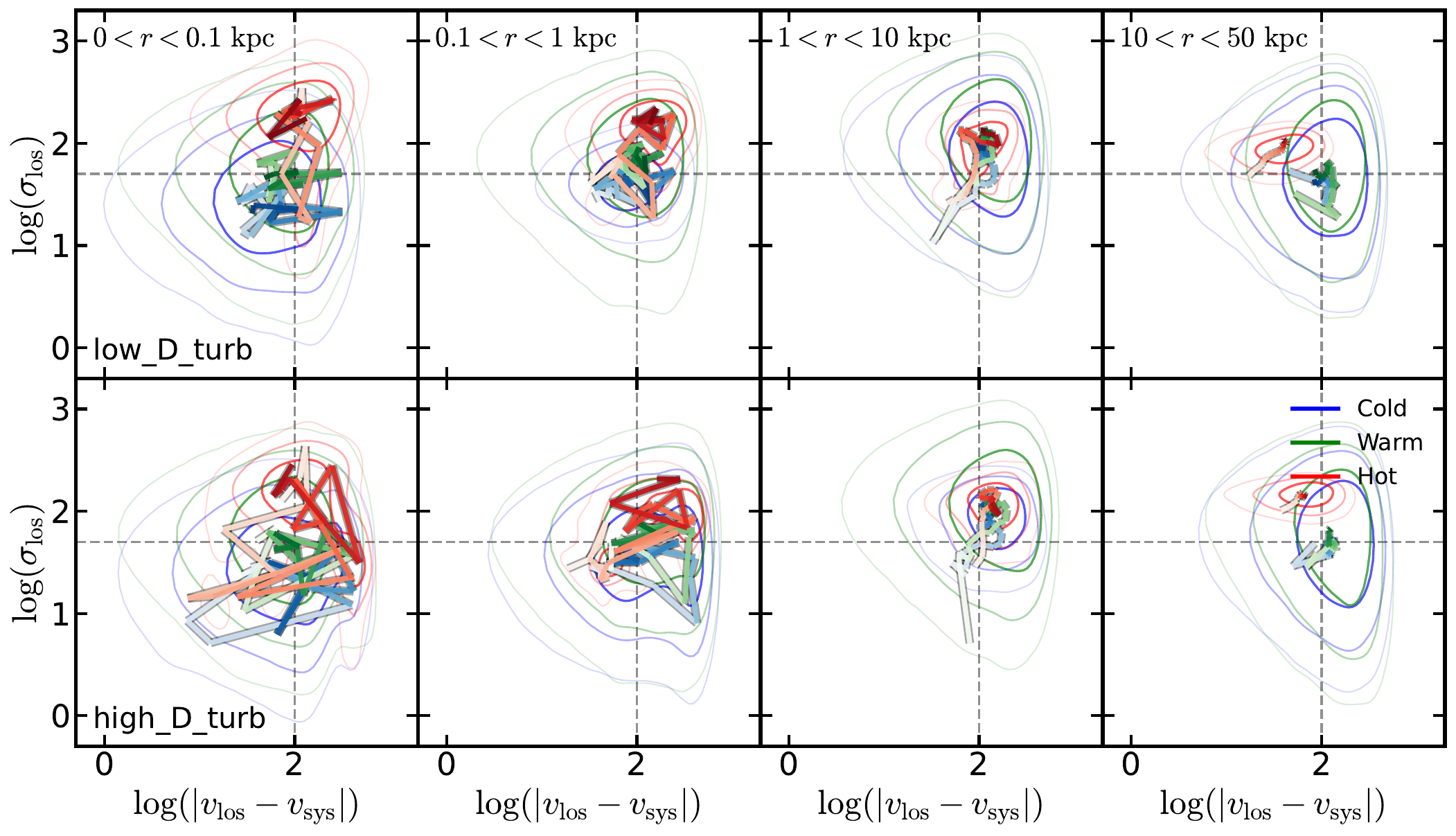}
\caption{Projected kinematic diagnostic (k-plot) for the continuously driven runs. The inner bins span a broader dynamic range in both velocity offset and line-of-sight dispersion than in the interrupted suite, with especially strong phase overlap and scatter inside $r<1\,{\rm kpc}$. Contours enclose the 85th, 92nd, and 97th percentile regions. Grey dashed lines mark the reference thresholds \(|v_{\rm los}-v_{\rm sys}|=100\,{\rm km\,s^{-1}}\) and \(\sigma_{\rm los}=50\,{\rm km\,s^{-1}}\).}
\label{kplot_driven}
\end{figure*}

\begin{figure*}
\centering
\includegraphics[width=0.95\textwidth]{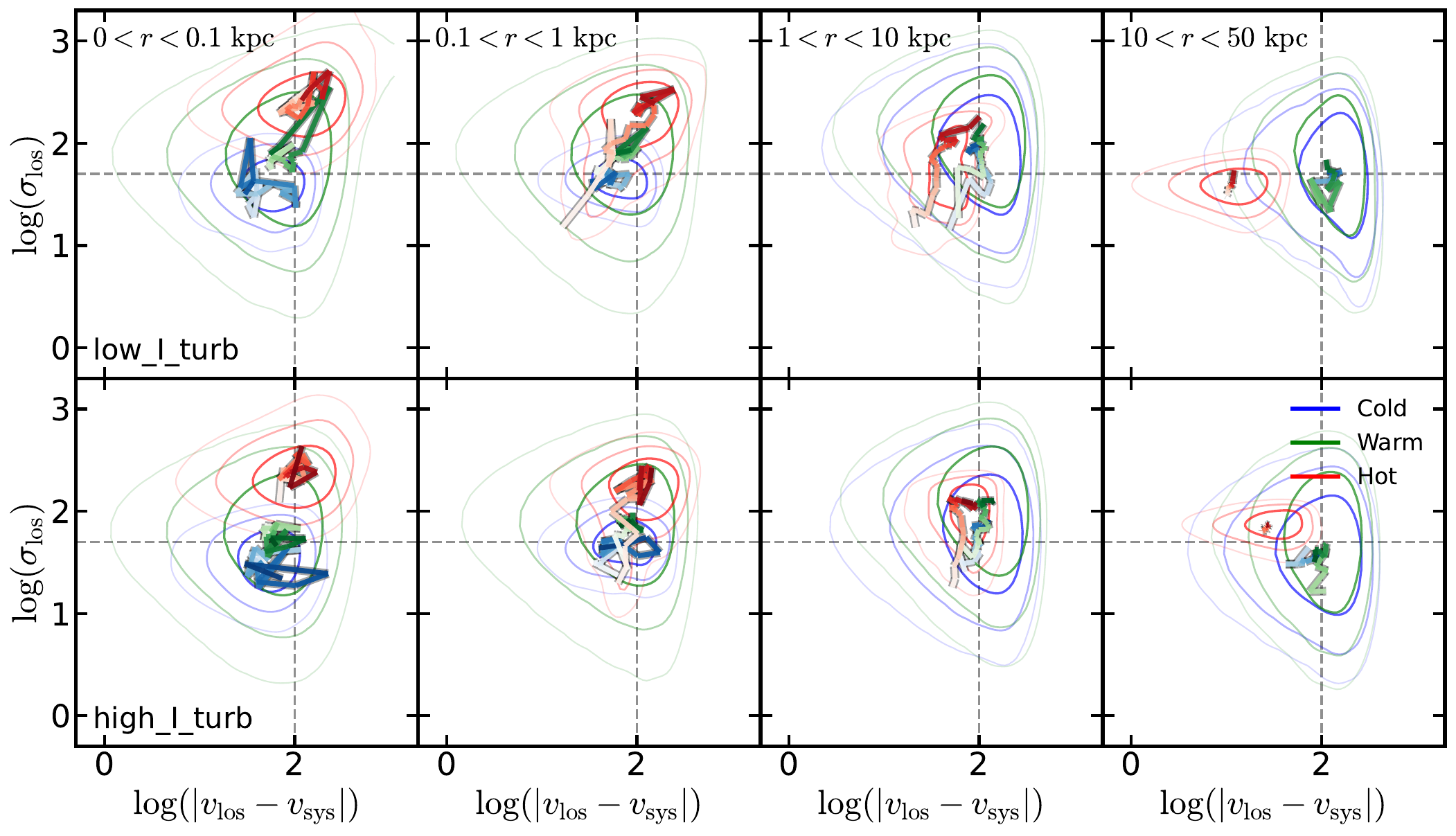}
\caption{Projected kinematic diagnostic (k-plot) for the interrupted-turbulence runs. Compared with the continuously driven suite, the loci are more compact and more phase-ordered, especially outside the meso-scale region. Contours enclose the 85th, 92nd, and 97th percentile regions. Grey dashed lines mark the reference thresholds \(|v_{\rm los}-v_{\rm sys}|=100\,{\rm km\,s^{-1}}\) and \(\sigma_{\rm los}=50\,{\rm km\,s^{-1}}\)}
\label{kplot_interr}
\end{figure*}

\subsubsection{k-plots}
\label{sec:results_kplots}

Figures~\ref{kplot_driven} and \ref{kplot_interr} provide the most direct projected kinematic discriminator between the DT and IT suites. We show the evolving k-plot diagnostic \citep{gaspari2018,Maccagni2021}, separating cold, warm, and hot gas in radial bins. For each snapshot, radial shell, and temperature phase, we project the gas along the \(z\)-axis and compute, in each projected pixel, the density-weighted line-of-sight centroid velocity and velocity dispersion. Since the simulations are analysed in the box frame, we set \(v_{\rm sys}=0\) and place each pixel in the \(\log_{10}|v_{\rm los}-v_{\rm sys}|\) versus \(\log_{10}\sigma_{\rm los}\) plane. The contours enclose the 85th, 92nd, and 97th percentile regions, while the coloured tracks mark the time evolution of the bulk of each phase from light to dark.

The reference thresholds \(|v_{\mathrm{los}}-v_{\mathrm{sys}}|=100~\mathrm{km\,s^{-1}}\) and \(\sigma_{\mathrm{los}}=50~\mathrm{km\,s^{-1}}\) separate the main projected kinematic regimes. The lower-left region traces relatively quiescent gas; the upper-right region identifies gas with both large bulk offsets and high internal dispersion; the upper-left region corresponds to turbulent or multi-component gas with modest centroid motion; and the lower-right region traces coherent high-velocity components, as commonly identified in absorption-line studies \citep[e.g.][]{Tremblay2016}. The location and overlap of the different thermal phases in this plane therefore provide a compact view of how tightly the multiphase gas is dynamically coupled.

The phase ordering is itself a CCA signature. The hot gas generally occupies the higher-dispersion part of the diagram, the cold gas tends to lie at lower dispersion, and the warm gas often bridges the two, as expected for material forming in turbulent mixing layers, conductive interfaces, shocks, or cooling wakes around cold filaments and clouds. The k-plots therefore do more than measure velocity/line broadening: they show how tightly the thermal phases are coupled in phase space. A compact, ordered locus indicates a more coherent precipitation channel, while broad and overlapping loci indicate stronger circulation, mixing, uplift/fallback, and multiphase coupling.

The strongest DT/IT contrast appears on meso scales, especially at \(0.1<r<1\,{\rm kpc}\). The DT suite (especially in \texttt{high\_D\_turb}) shows a more dispersed time trajectory, stronger cold--warm--hot overlap, and decreased migration toward the diagonal \(\sigma_{\rm los}\simeq |v_{\rm los}-v_{\rm sys}|\), given the higher stochasticity of the continuous stirring. This indicates that persistent stirring converts the inner halo into a circulation-dominated multiphase medium in which condensation, jet-driven uplift, fallback, turbulent mixing, and cloud--halo interactions coexist. In observable terms, the DT suite predicts broad and overlapping multiphase velocity distributions, with strong kinematic coupling between the cold molecular gas traced by CO, the warm ionized gas traced by H\(\alpha\), and the hot X-ray emitting atmosphere. This behaviour is the projected counterpart of the CCA condensation cascade, in which cold and warm filaments condense out of the turbulent hot halo and retain correlated ensemble kinematics across phases, as found in several multi-wavelength observations \citep[e.g.][]{gaspari2018,Tremblay2018,Maccagni2021,Temi2022,Olivares2022}.

The IT suite, by contrast, contracts toward more compact and phase-ordered loci, especially outside the central bin. The cold and warm phases remain less blended with the hot component and show less turbulent broadening at fixed velocity offset, consistent with a more connected rainy state in which condensed gas is less thoroughly mixed before reaching the nucleus. Observationally, this would correspond to narrower and more phase-separated velocity structures, with cold/warm gas more clearly organized into ordered inflow, precipitation, or filamentary rain, and with weaker overlap between the CO, H\(\alpha\), and X-ray kinematic distributions. In the outermost bin, \(10<r<50\,{\rm kpc}\), the visible distribution is dominated mainly by the hot component, particularly in \texttt{low\_I\_turb}; this suggests that cold/warm condensation becomes more centrally concentrated rather than absent from the system altogether.

Together, the three diagnostics support a consistent interpretation of CCA. The \(C\)-ratio identifies where the hot atmosphere becomes susceptible to precipitation, \(\mathrm{Ta_t}\) shows that the resulting multiphase flow remains dispersion-dominated rather than disc-dominated, and the k-plots show how turbulence controls the phase-space coupling of hot, warm, and cold gas. Persistent stirring therefore changes how condensation is transmitted across scales. In the DT suite, the meso-scale flow behaves more like a turbulent circulation that filters and decorrelates mass and angular momentum before they reach the SMBH, leaving broad, coupled multiphase kinematics as the observable imprint. In the IT suite, the same CCA condensation cascade more often remains organized into connected rain, producing stronger sink feeding, higher torque coherence, faster effective jet-axis reorientation, and narrower, more ordered multiphase kinematics.

\section{Discussion}
\label{sec:discussion}

The main result of this work is that SMBH spin reorientation in CCA is controlled not by the instantaneous accretion rate alone, but by the coherent vector delivery of mass and angular momentum through the meso-scale multiphase bridge.
Persistent stirring -- i.e.\ driven turbulence -- changes whether cold and warm gas from larger scales remain radially connected to the nucleus and whether successive angular-momentum increments add constructively or cancel before reaching the SMBH. In the DT suite, the sink can still receive stochastic gas inflow, but the inflow is fragmented, directionally variable, and cancellation-prone, so the spin drifts slowly. In the IT suite, decaying turbulence allows the same chaotic cold rain to remain connected for longer intervals, producing a higher \(\dot{M}_{\rm sink}\), higher \(\chi_j\), and faster effective jet-axis reorientation.

\begin{figure*}[!t]
    \centering
    \includegraphics[width=0.75\textwidth]{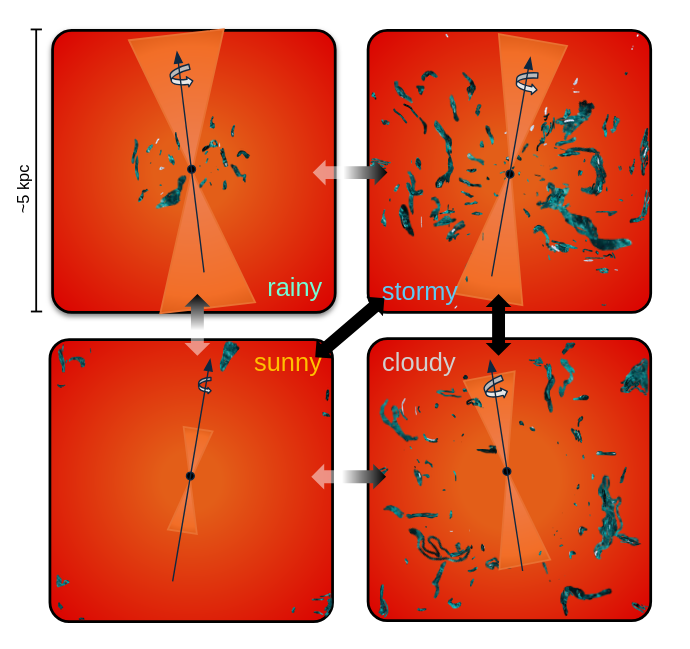}
    \vspace{-0.1cm}
    \caption{Schematic diagram of the spin-regulated {\sc BlackHoleWeather} cycle. The four weather states \sunny, \stormy, \cloudy, and \rainy\ describe how the multiphase halo delivers mass and angular momentum to the SMBH, thereby setting the spin response and the orientation of the next jet episode. \rainy\ states maintain a connected cold/warm channel to the sink, favouring high \(\dot{M}_{\rm sink}\), coherent torque delivery, and stronger secular spin evolution. \stormy\ states contain extended filamentary precipitation and a broad hot--warm--cold bridge, producing bursty fueling and rapidly varying misaligned torques with limited long-term coherence. \cloudy\ states retain fragmented or jet-processed multiphase gas, but central delivery is intermittent and torque cancellation is efficient. \sunny\ states are hot-dominated or feedback-cleared central configurations with weak sink feeding and a torque-starved SMBH. Curved arrows indicate the vector feedback loop: precipitation feeds the SMBH, coherent torques update the spin, the spin sets jet power and direction, and feedback clears, mixes, or reconnects the next multiphase episode.}
    \label{fig:weather_cycle_cartoon}
\end{figure*}

Following the companion {\sc BlackHoleWeather} studies (B26a,b; C26a,b; \citealt{gaspari2020}), we use \emph{weather states} as operational labels for the scale-dependent configuration of the hot atmosphere and its multiphase condensate. In the present spin-coupled problem, these labels are most useful when interpreted through the central meso-scale bridge connecting halo condensation to sink feeding. A \emph{rainy} state is a connected precipitation state: cold and warm gas reach the sink, sustain enhanced accretion, and can deliver angular momentum coherently. A \emph{stormy} state is extended, filamentary, and burst dominated, with a broad hot--warm--cold bridge across meso-to-macro radii; multiphase gas is abundant, but the angular-momentum supply varies rapidly, so instantaneous misaligned torques can be large, while secular coherence remains limited. The \emph{cloudy} state contains fragmented or jet-processed multiphase gas with intermittent or inefficient nuclear delivery. A \emph{sunny} state is hot-dominated or feedback-cleared in the centre: sink feeding is weak, the SMBH is torque-starved, and any remaining cold gas is mostly displaced to larger radii. These categories are scale dependent and can overlap, e.g. the central $\sim 10-100$ pc may be sunny while the inner kpc remains stormy or cloudy.

In the spin-coupled runs, the weather state determines not only whether cold/warm gas is present but also whether it reaches the sink with a coherent angular-momentum direction. High-accretion rainy states are therefore coherent torque-delivery states, capable of reorienting the SMBH. Stormy states contain more extended, turbulent, and bursty precipitation, often with abundant gas at \(r>1\,{\rm kpc}\), but the delivered torque can decorrelate rapidly. Cloudy or sunny states may still contain cold gas at meso or halo radii, but the nucleus is temporarily decoupled from that reservoir or is fed through fragmented threads whose angular momenta cancel.

Figure~\ref{fig:weather_cycle_cartoon} summarizes this spin-coupled weather-cycle interpretation. The red background denotes the hot atmosphere, the cyan structures trace condensed multiphase gas, the orange cones mark the spin-aligned jet direction, and the central axis/curved arrow represents the SMBH spin and its effective reorientation. The diagram should not be read as a deterministic chronological loop, but as a schematic phase-space map of the main routes between connected rain, turbulent circulation, fragmentation, and feedback clearing. The fading arrows indicate qualitatively favored transitions, with lower opacity corresponding to less frequent or less direct paths. Consistently with Figure~\ref{inflow_map_cold_main}, the driven-turbulence runs tend to follow a fragmentation-dominated sequence, \emph{rainy} \(\Rightarrow\) \emph{stormy} \(\Rightarrow\) \emph{sunny} \(\Rightarrow\) \emph{stormy}/\emph{cloudy}, in which persistent stirring repeatedly breaks the cold/warm bridge to the sink. The interrupted-turbulence runs do not enter the same prolonged fragmentation-dominated \emph{sunny} state. Their central clearing is instead more directly tied to strong post-rain feedback episodes: the nucleus can become temporarily hot and torque-starved, but the cold/warm bridge reconnects more readily once the feedback episode fades.

This diagram can be read as the spin-coupled extension of the weather-cycle pictures developed in B26a and C26a. In B26a,b, where no explicit jet is present, the main control parameter is the level of turbulent stirring: weak turbulence favors compact, centrally retained \emph{rainy} condensation, whereas stronger turbulence delays condensation and spreads the cold/warm phase into an extended, filament-rich \emph{stormy} configuration. C26a,b add fixed-axis mechanical feedback, showing that jets do not erase this turbulence-regulated dichotomy, but reshape it through compression, uplift, entrainment, and mixing: the low-turbulence case remains closer to a centrally retained \emph{rainy} cycle, while the high-turbulence case evolves from an extended \emph{stormy} phase toward a more \emph{cloudy} state with inefficient central delivery. The key difference here is that the jet direction is no longer fixed. The same weather loop therefore acquires a vector component: feedback geometry can redistribute energy over a broader solid angle, produce more extended central clearing, and convert a scalar change in accretion state into a change in torque coherence and SMBH spin orientation.

Through this coupling between the feeding bridge, SMBH spin, and jet geometry, the weather state also controls the time-domain and observable signatures of jet-axis reorientation. Rainy states preserve a high \(\dot{M}_{\rm sink}\), high \(\chi_j\), and coherent \(j_\parallel/j_\perp\) structure, producing faster effective jet-axis reorientation and enhanced low-frequency accretion power. Cloudy, stormy, or continuously stirred states can contain abundant multiphase gas, but their angular-momentum delivery is fragmented, sign-changing, and cancellation-dominated, producing slower spin drift, stronger high-frequency damping, and broader, more overlapping multiphase kinematic loci.

Within the controlled scope of the present experiment, the \textit{Hybrid} model connects the resolved pc-scale inflow to the SMBH through an ISCO-based subgrid closure, and the Blandford--Znajek prescription keeps the magnetic-flux normalization fixed. These choices isolate the meso-scale torque-coherence pathway from additional uncertainties in magnetic-flux accumulation and unresolved disc alignment. Future work should couple this pathway to self-consistent magnetic-flux evolution, sub-pc alignment physics, and emissivity-weighted synthetic CO, H\(\alpha\), and X-ray observables.

 \subsection{Comparison with previous work}
 \label{sec:comparison}

P26a placed the \textit{Hybrid} spin prescription in the broader context of semi-analytic spin models, sub-grid accretion treatments, prescribed jet reorientation, and relativistic spin closures. We do not repeat that full comparison here. Instead, we use the validated \textit{Hybrid} closure to isolate the specific new ingredient of this work: how the resolved CCA weather state regulates the coherence, variability, and observable imprint of the torque delivered to the SMBH.

The closest non-CCA connection is with chaotic or episodic accretion models, in which the long-term spin history depends on the coherence of successive accretion/merger events rather than on the mass budget alone \citep{volonteri2005,king2006,king2008,perego2009,dotti2013}. Our results support this picture, but replace imposed cosmological/merger episodes with resolved multiphase weather. In the present simulations, the torque episodes emerge from condensation, cloud--cloud interactions, fallback, uplift, and turbulent mixing. The quantities \(\dot{M}_{\rm sink}\), \(\chi_j\), \(j_\parallel\), and \(j_\perp\) therefore measure how halo-scale CCA is converted into SMBH spin response.
Simulations with spin-dependent or imposed reorienting jets have shown that changing the jet direction can alter cavity morphology, heating anisotropy, and long-term gas regulation \citep{dubois2014,cielo2018,horton2020,talbot2021,talbot2022,beckmann2019}, consistent with observed misaligned cavities and bent or S/Z-shaped radio structures \citep{krause2019,bruni2021,ubertosi2023,misra2025}. 

However, from the observational side, SMBH spin in AGN is hard to constrain. Mostly, it is done through relativistic X-ray reflection spectroscopy, where the shape of the broad Fe~K$\alpha$ line, the associated Fe~K edge, and the Compton hump encode the location of the innermost stable circular orbit and hence the black-hole spin \citep[e.g.][and references therein]{brenneman13,reynolds21,bambi21,siskreynes26}. Current compilations of reflection-based spin measurements suggest a population biased toward high prograde spins, but this distribution should be interpreted with caution because the available sample is heterogeneous and affected by substantial statistical and systematic uncertainties \citep{siskreynes26}. In practice, spin is often degenerate with other parameters such as disc inclination, emissivity profile, ionization structure, absorption, and assumptions about the inner disc radius. In some analyses it is fixed or only weakly constrained \citep[e.g.,][]{serafinelli23}, which can artificially favour boundary values such as $a_\ast\simeq+0.998$ in distribution papers. Limited CCD spectral resolution, especially in complex absorbed AGN, has historically made it difficult to separate broad relativistic reflection from narrow emission/absorption components. New XRISM observations of MCG--6-30-15 show that high-resolution spectroscopy can isolate such narrow features and may help recover less biased spin constraints, although broadband, time-resolved modelling remains essential to break the remaining degeneracies \citep{brenneman25}.

The main physical lineage of this work is the CCA and \textsc{BlackHoleWeather} framework. Earlier CCA studies showed that turbulence, cooling, cloud condensation, collisions, and AGN heating form a recurrent multiphase cycle in which cold gas rains toward the SMBH and boosts feeding relative to smooth hot-mode capture \citep{gaspari2013,gaspari2015,gaspari2017,gaspari2020}. Here we extend that framework from scalar mass feeding to vector torque delivery: the weather state determines not only how much gas reaches the sink, but also whether its angular momentum remains coherent long enough to reorient the SMBH.

In this work we presented how the inclusion of spin-jet coupling can affect the observational CCA diagnostics. The condensation ratio, turbulent Taylor number, and k-plot were introduced to connect precipitation, turbulence/rotation balance, and multiphase kinematic coupling in hot halos \citep{gaspari2018}. Observational applications have shown that CO, H\(\alpha\), and X-ray gas often display linked but phase-dependent kinematics, consistent with condensation from a turbulent hot atmosphere \citep{Tremblay2018,Maccagni2021,Temi2022,Olivares2022,Wang2023}. Here we link these diagnostics to spin response: IT runs preserve connected rainy intervals with high \(\dot{M}_{\rm sink}\) and high \(\chi_j\), producing faster effective reorientation and enhanced low-frequency accretion power; DT runs fragment the multiphase bridge, increase angular-momentum cancellation, steepen the high-frequency PSD damping, and produce broader, more overlapping multiphase kinematic loci.

\section{Conclusions}
\label{sec:conclusions}

Building on the chaotic cold accretion framework and on the \textsc{BlackHoleWeather} companion sequence (B26a,b, C26a,b), we used the \textit{Hybrid} spin-evolution model validated in P26a to study how weather-regulated feeding controls the response of SMBH spin, effective jet-axis reorientation, accretion variability, and multiphase halo kinematics. The controlled experiment is the persistence of large-scale turbulent stirring after cooling and jet feedback are activated: in the DT suite turbulence remains continuously driven, whereas in the matched IT suite the same initial turbulent field is allowed to decay. Our key conclusion is that persistent stirring mainly disrupts the radial mass and angular-momentum continuity of the meso-scale CCA bridge, reducing the coupling between sink feeding and coherent torque delivery without suppressing large-scale condensation. The main results are:

\begin{itemize}

\item Persistent stirring breaks nuclear connectivity rather than large-scale condensation. At large radii, all four runs show comparable gross inward fluxes of order \(\sim10^2\,M_\odot\,{\rm yr^{-1}}\). The DT/IT split appears inside the central kiloparsec and becomes strongest near the sink: by \(r\sim10\,{\rm pc}\), the IT runs retain \(\sim0.3\)--\(2\,M_\odot\,{\rm yr^{-1}}\), whereas the DT runs drop to \(\sim10^{-3}\)--\(10^{-2}\,M_\odot\,{\rm yr^{-1}}\). Persistent stirring therefore disrupts the meso-scale bridge that transmits condensed gas from the halo to the SMBH.

\item Spin-coupled CCA is a vector feeding problem. The relevant quantity is not only how much cold gas reaches the SMBH but also whether that gas reaches the sink with a coherent angular-momentum direction. Rainy states correspond to connected cold/warm channels with enhanced sink feeding and coherent torque delivery. Cloudy, stormy, or sunny states can still contain multiphase gas, but the nucleus is weakly connected, recently cleared, or fed by fragmented threads whose angular momenta cancel.

\item Torque coherence is the sink-scale control parameter. Strongly misaligned accretion episodes, with large \(j_\perp\), occur in all runs, but misalignment alone is not sufficient to drive sustained reorientation. Rapid spin-axis changes require high \(\dot{M}_{\rm sink}\), large perpendicular torque, and a coherence time long enough for successive torque increments to be added constructively. The IT runs, especially \texttt{low\_I\_turb}, maintain the highest coherence and accretion duty cycles, while the DT suite more often decouples coherent torque delivery from sustained sink feeding, particularly in the high-turbulence driven case.

\item Jet-axis reorientation is an emergent weather-regulated response, not an imposed fixed precession period. In this low-spin suite, the SMBH responds mainly through orientation changes rather than through large sustained changes in spin magnitude. After the activation time, the DT runs mostly settle to effective reorientation rates of \(\sim10^{-4}\)--\(10^{-3}\,{\rm deg\,Myr^{-1}}\), whereas the IT runs typically remain at \(\sim10^{-2}\)--\(10^{-1}\,{\rm deg\,Myr^{-1}}\) and can briefly reach a few \({\rm deg\,Myr^{-1}}\) during coherent retrograde episodes. The near-null episode in \texttt{low\_I\_turb} is best interpreted as a spin reset driven by coherent retrograde feeding, rather than as smooth periodic precession.

\item Accretion variability carries the temporal imprint of connected rain. All runs retain broad-band, flicker-like CCA variability in the power spectra. The IT suite has enhanced low-frequency power, especially in \texttt{low\_I\_turb}, consistent with longer connected rainy episodes. The DT suite shows steeper high-frequency damping, consistent with persistent stirring filtering cloud-scale fluctuations through fragmentation, mixing, and angular-momentum cancellation before they reach the sink. The fitted breaks cluster near \(f_{\rm b}\sim20\,{\rm Myr^{-1}}\), corresponding to \(\sim0.05\,{\rm Myr}\), and are best interpreted as a common inner response scale rather than the timescale of the multi-Myr weather cycle.

\item The halo remains CCA-like, but its multiphase kinematic imprint changes. The \(C\)-ratio shows that both suites can enter the precipitation regime, with the soft-X phase tracing where the hot atmosphere becomes susceptible to cooling. The turbulent Taylor number remains mostly \(\mathrm{Ta_t}\lesssim1\), confirming that the multiphase flow is dispersion-dominated rather than rotationally supported. The projected k-plots show that the DT suite produces broader and more overlapping cold/warm/hot loci, especially inside the central kpc, while the IT suite produces narrower, more phase-ordered structures consistent with connected precipitation. In observable terms, persistent stirring predicts stronger kinematic coupling between CO-like molecular gas, H\(\alpha\)-like warm gas, and X-ray-emitting plasma.

\end{itemize}

Overall, these simulations extend the \textsc{BlackHoleWeather} framework from scalar mass feeding to 3D angular-momentum delivery. In CCA, the same multiphase feeding cycle that regulates when the SMBH is fed also determines whether the delivered angular momentum is coherent enough to reorient the spin and redirect the jet. Future work will expand the simulated feeding regimes and couple this meso-scale torque-coherence pathway to self-consistent magnetic-flux evolution, with the purpose to provide physically-motivated prescriptions that could help cosmological simulations and semi-analytic models to study population-wide SMBH spin statistics.\\

\begin{acknowledgements}
The {\sc BlackHoleWeather} authors acknowledge key funding support from the European Research Council (ERC) under the European Union's Horizon Europe research and innovation programme (Consolidator Grant BlackHoleWeather, No.~101086804; PI: Gaspari). Views and opinions expressed are, however, those of the author(s) only and do not necessarily reflect those of the European Union or the European Research Council Executive Agency; neither the European Union nor the granting authority can be held responsible for them.
We acknowledge ISCRA for awarding this project access to the LEONARDO supercomputer, owned by the EuroHPC Joint Undertaking, hosted by CINECA (Italy).
We acknowledge EuroHPC Joint Undertaking for awarding us access to Jupiter at FZJ, Germany.
VO acknowledges support from the DICYT ESO-Chile Comite Mixto PS 1757, Fondecyt Regular 1251702, and CIRAS-AI Project, code FIUF137139-USACH.
PT acknowledges support from NASA NNH22ZDA001N Astrophysics Data and Analysis Program under award 24-ADAP24-0011. RS acknowledges funding from the CAS-ANID grant No.~220016.
FMM acknowledges support from the Next Generation EU funds within the PNRR, Mission 4 -- Education and Research, Component 2 -- From Research to Business (M4C2), Investment Line 3.1 -- Strengthening and creation of Research Infrastructures (project IR0000034 ``STILES'').
We thank Hung-Yi Pu, Martin Fournier, and Amirnezam Amiri for comments and discussions.
We thank the organizers and participants of the following conferences for the stimulating discussions that helped improve this work: `BlackHoleWeather I' (Sexten, ITA), `Massive Black Hole Spin' (Edinburgh, UK).
\end{acknowledgements}

\bibliography{lib}{}
\bibliographystyle{aa}
\clearpage
\onecolumn
\appendix
\section{Profiles}\label{app:profiles}
We show here the time evolution of the radial profiles of the thermodynamic properties of our simulations. In particular, we note that in the driven runs (DT) continuous turbulence injection prevents density and pressure from building up in the nucleus. 
Compared with the fixed-axis jet case of C26a (their Figure~7), the cold/molecular component is lower by \(\sim1\) dex in the central region (\(r<100\,{\rm pc}\)), while the warmer and hot phases remain broadly comparable. This suggests that spin-driven jet-axis reorientation may help disrupt, redistribute, or reheat dense cold gas, reducing the formation of a very compact nuclear reservoir.
\begin{figure*}[h]
\includegraphics[width=\textwidth]{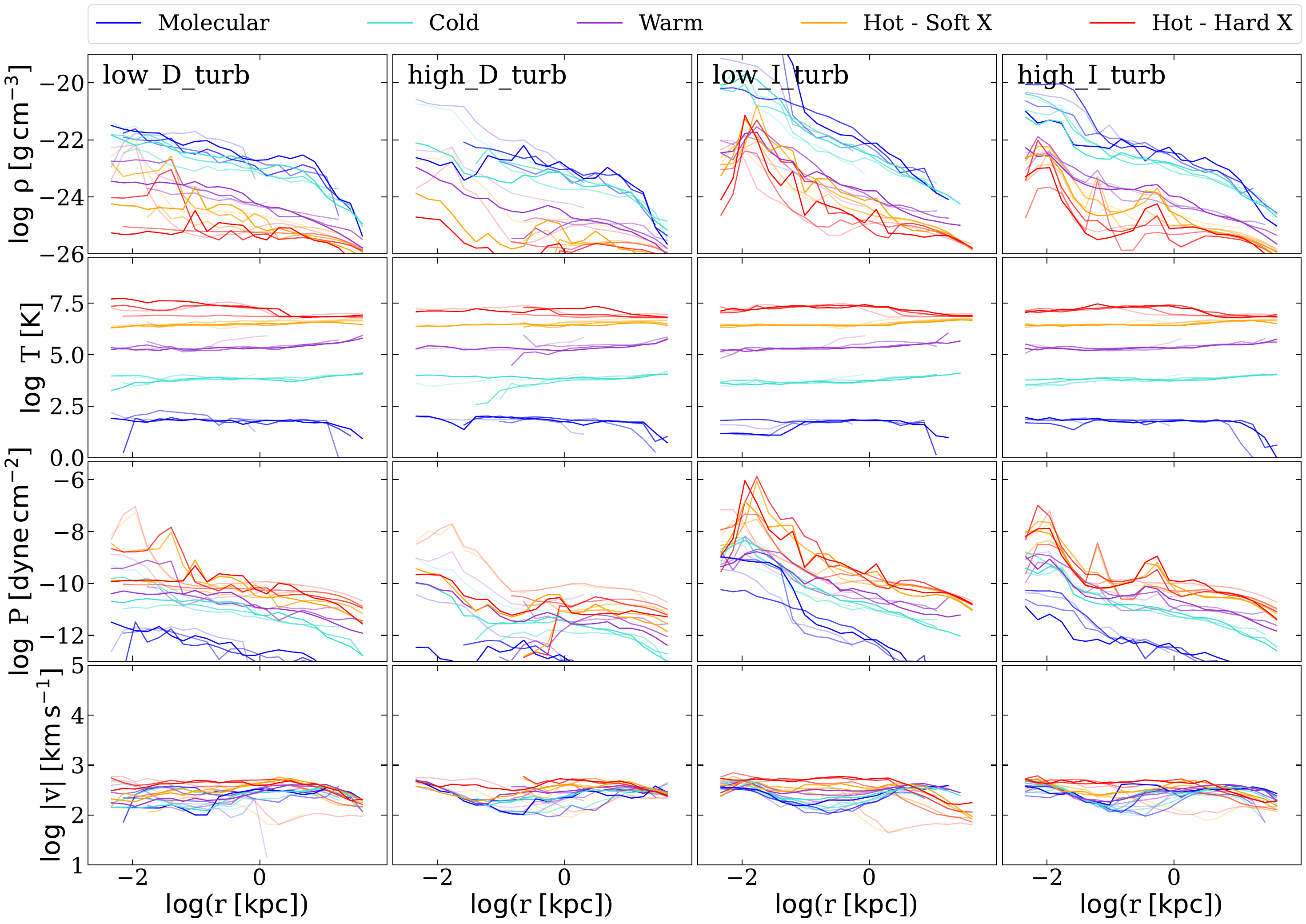}
\caption{Radial profiles of the density, temperature, pressure and velocity for all runs and for different phases. Later times correspond to a higher degree of transparency.}
\label{profiles}
\end{figure*}

\section{Complementary phase-separated time--radius maps}\label{app:maps}

In the main text we show the maps that carry the argument most directly: the cold inflow maps (Figure~\ref{inflow_map_cold_main}), and the hot outflow maps (Figure~\ref{outflow_map_hot_main}). 
These diagnostics are the most efficient representation of the SMBH weather cycle because they isolate, respectively, the global feeding continuity, the rainy cold channel, and the sunny feedback-clearing phase. The remaining maps are collected here because they corroborate the same cycle but are more diagnostic of the phase decomposition than of the core storyline.

\subsection{Inflow by phase}

The hot inflow map is the least discriminatory of the inflow diagnostics. As shown in Figure~\ref{inflow_map_hot}, the hot component remains comparatively smooth and quasi-volume filling across all runs, with typical inward speeds of order a few $10$ to a few $10^2\,{\rm km\,s^{-1}}$. The DT/IT differences are therefore milder here than in the condensed phases. This is precisely why the hot inflow is better kept in the appendix: it confirms the background feeding environment, but it does not isolate the rainy phases nearly as cleanly as the cold inflow does.

The warm inflow, by contrast, mirrors the cold phase and shows that the rainy channel is genuinely multiphase. In Figure~\ref{inflow_map_warm}, the DT runs again develop large interruptions at $r\sim0.05$--$1\,{\rm kpc}$ --- especially around $\tau\sim6$--8 in \texttt{low\_D\_turb} and $\tau\sim6$--12 in \texttt{high\_D\_turb} --- whereas the IT runs preserve a more continuous inward bridge once cooling begins. The warm inflow speeds are also broadly comparable to the cold-phase values, again emphasizing that the decisive contrast is the persistence of the inward connection rather than a strong change in the characteristic inflow speed.

\begin{figure*}[t]
\includegraphics[width=0.5\textwidth]{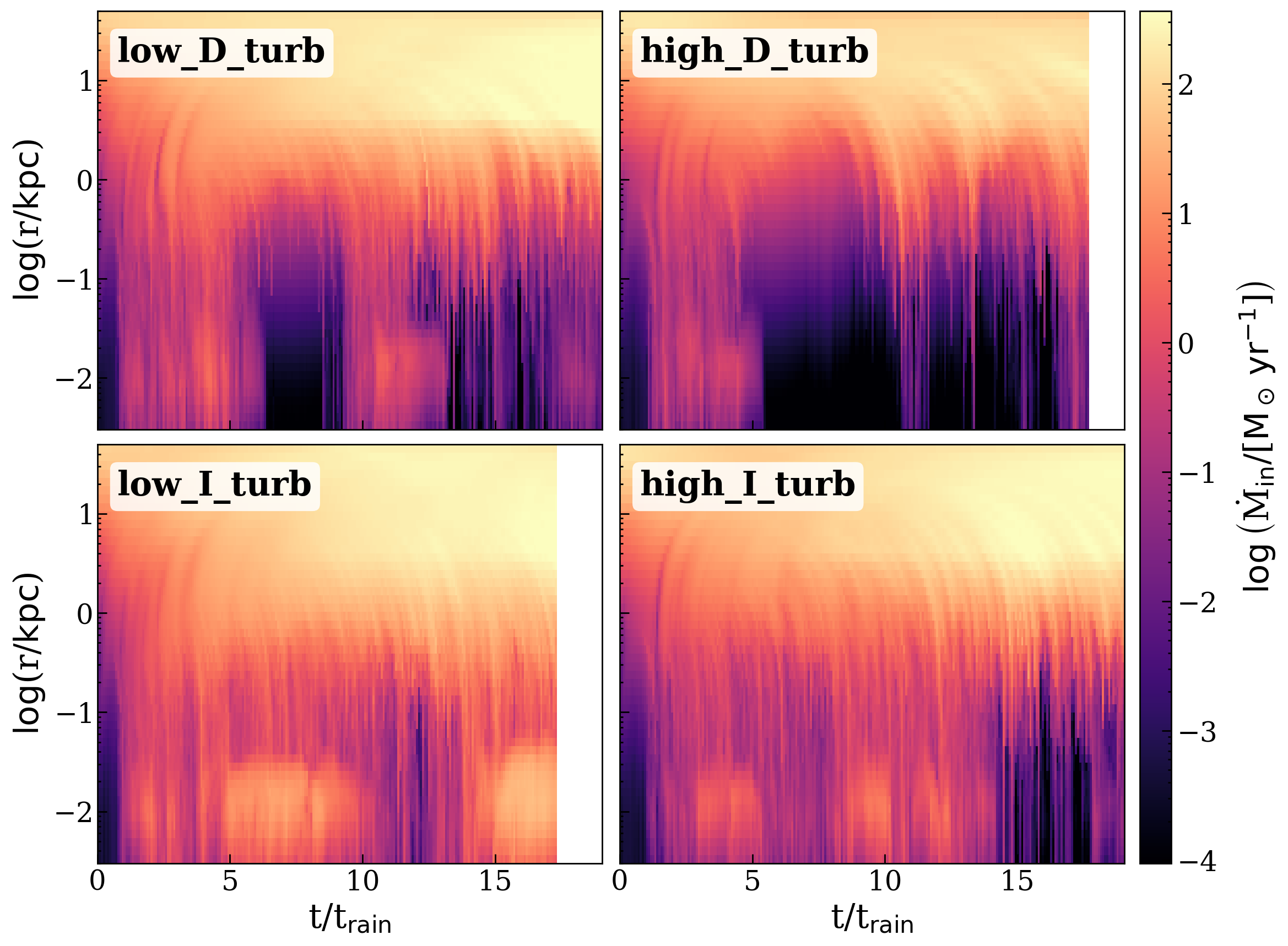}
\includegraphics[width=0.5\textwidth]{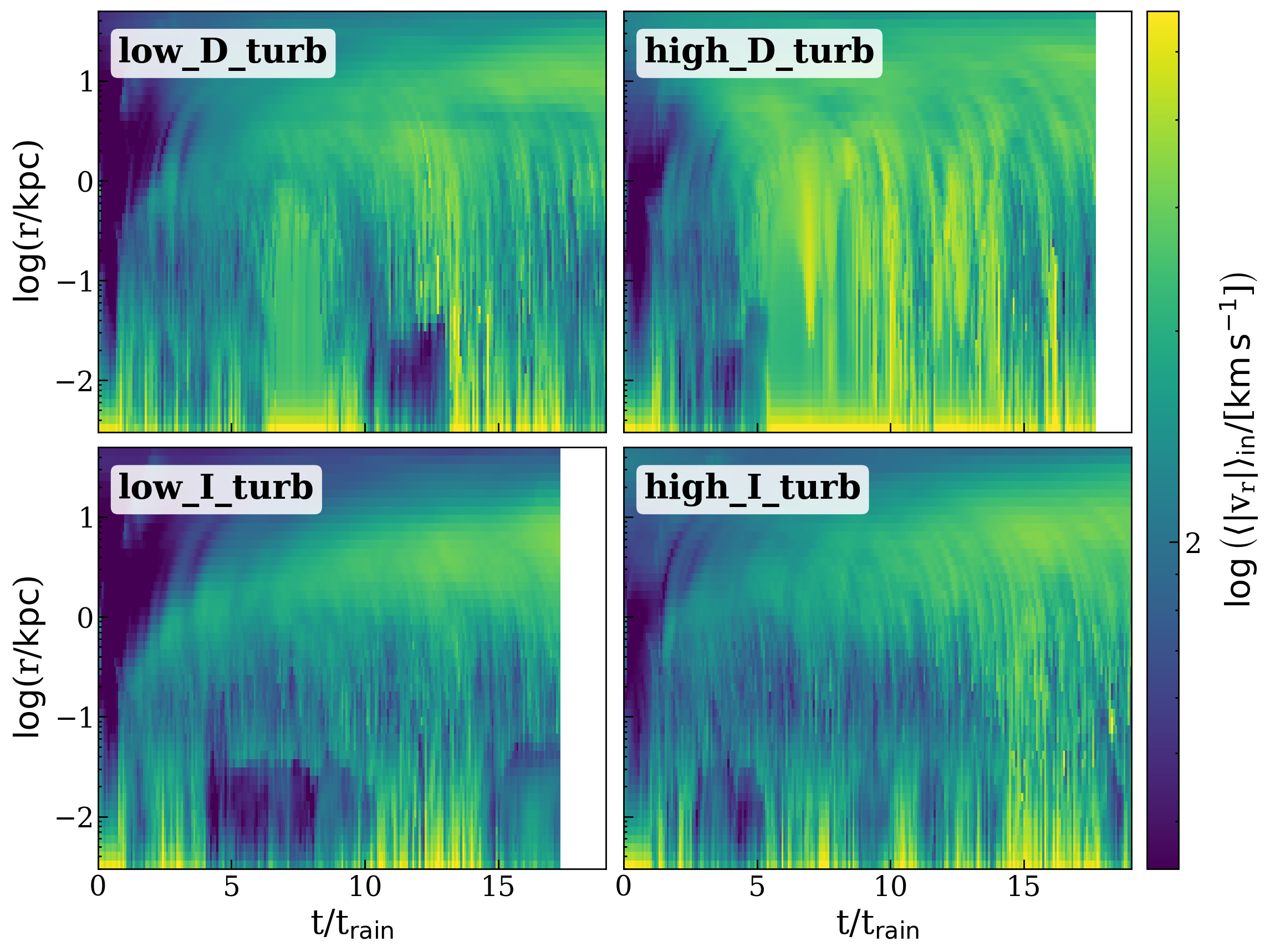}
\caption{Time--radius maps of the total inflow mass rate (top) and inflow radial velocity amplitude (bottom) for all four runs. The IT runs preserve a more continuous inflow bridge from $\sim1$--$10\,{\rm kpc}$ to the sink over broad intervals, while the DT runs show repeated breaks at $r\lesssim0.1$--$1\,{\rm kpc}$, strongest in \texttt{high\_D\_turb}.}
\label{inflow_map}
\end{figure*}

\begin{figure*}[t]
\includegraphics[width=0.5\textwidth]{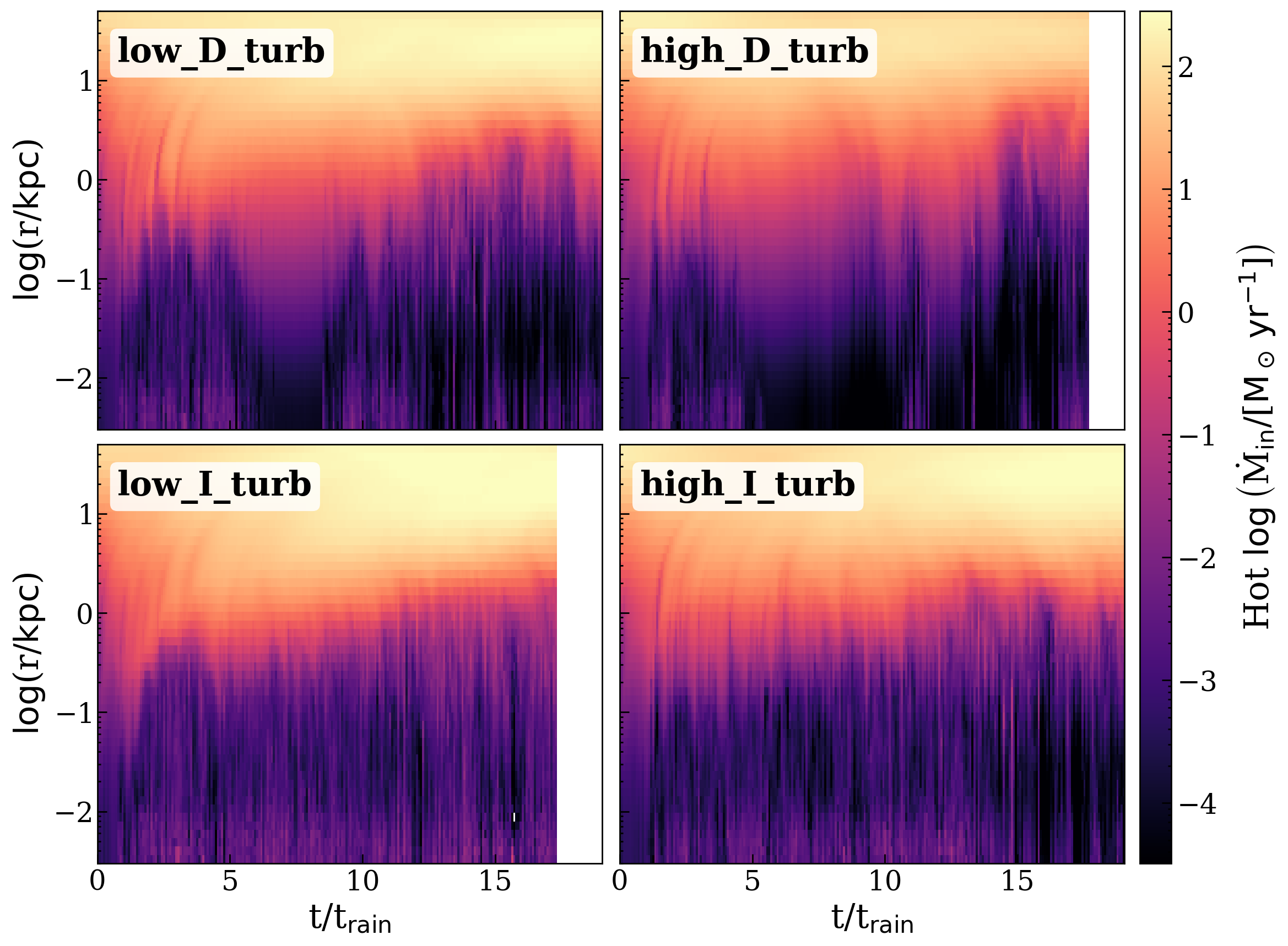}
\includegraphics[width=0.5\textwidth]{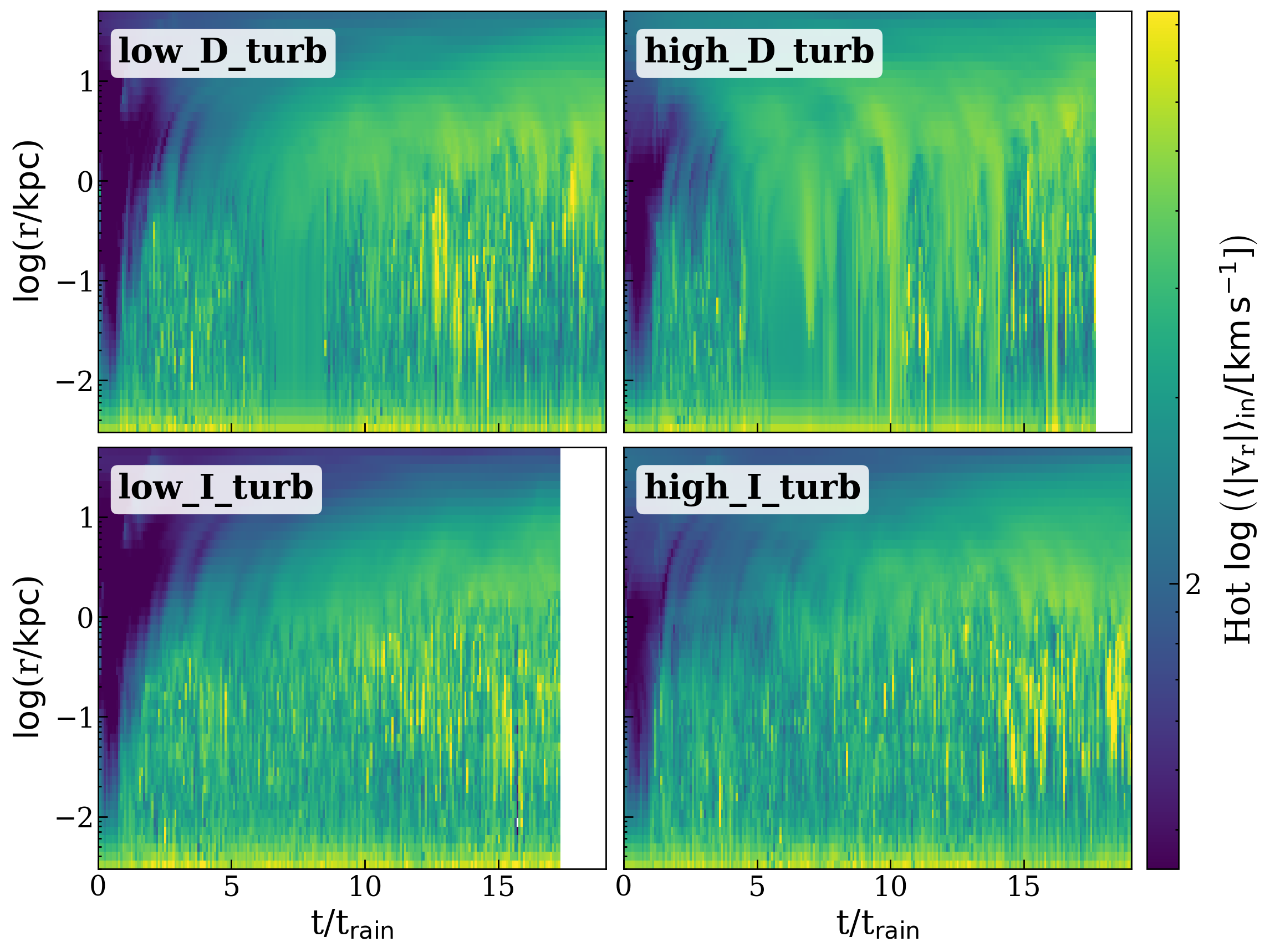}
\caption{Time--radius inflow mass-rate and radial-velocity maps for the hot phase. The hot inflow is comparatively smooth and volume filling, so the contrast between DT and IT is weaker here than in the condensed phases.}
\label{inflow_map_hot}
\end{figure*}

\begin{figure*}[t]
\includegraphics[width=0.5\textwidth]{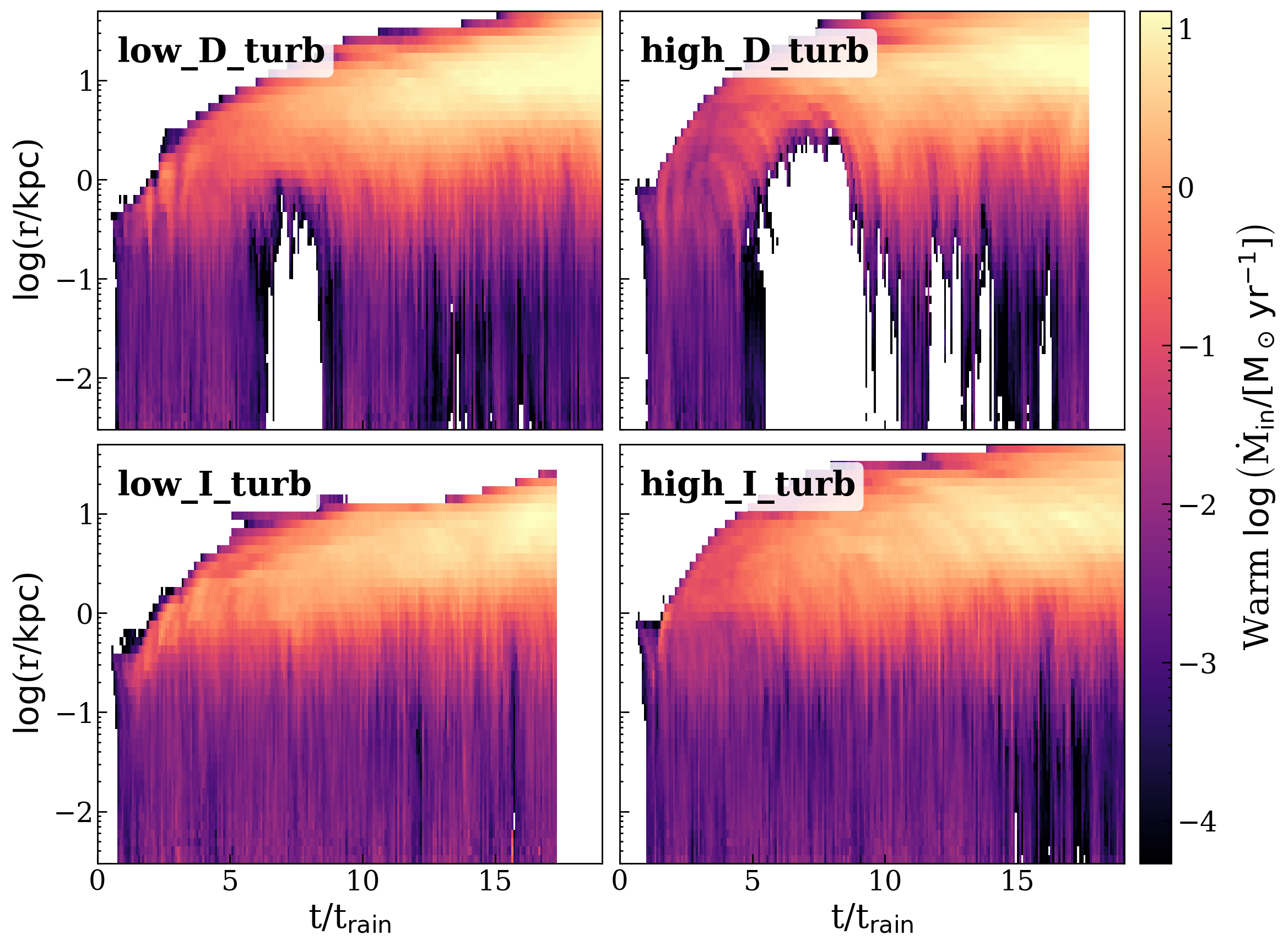}
\includegraphics[width=0.5\textwidth]{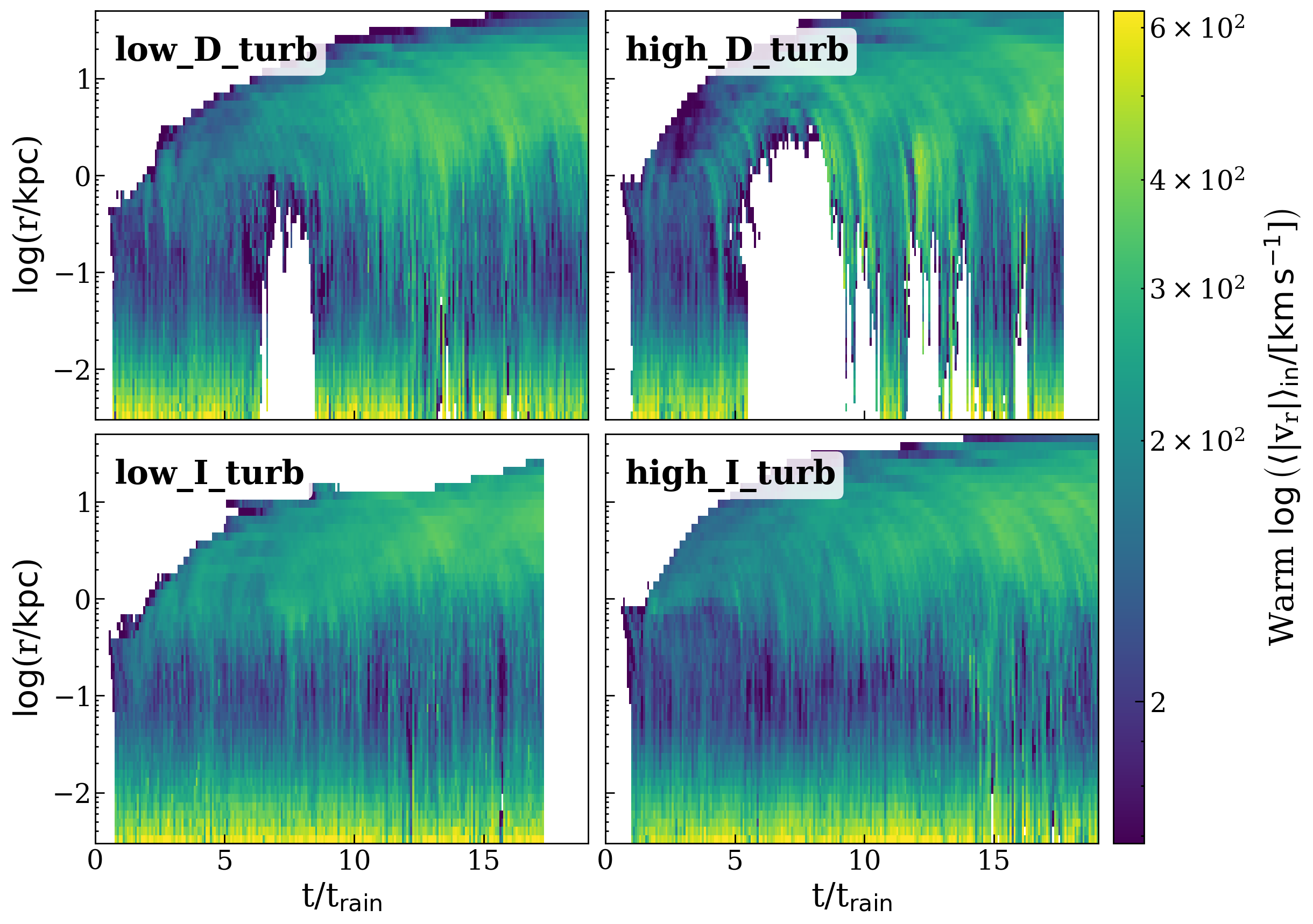}
\caption{Time--radius inflow mass-rate and radial-velocity maps for the warm phase. The warm component closely tracks the cold rain and shows the same DT/IT contrast in radial continuity, confirming that rainy weather is a genuinely multiphase phenomenon.}
\label{inflow_map_warm}
\end{figure*}

\subsection{Outflow by phase}

\begin{figure*}[t]
\includegraphics[width=0.5\textwidth]{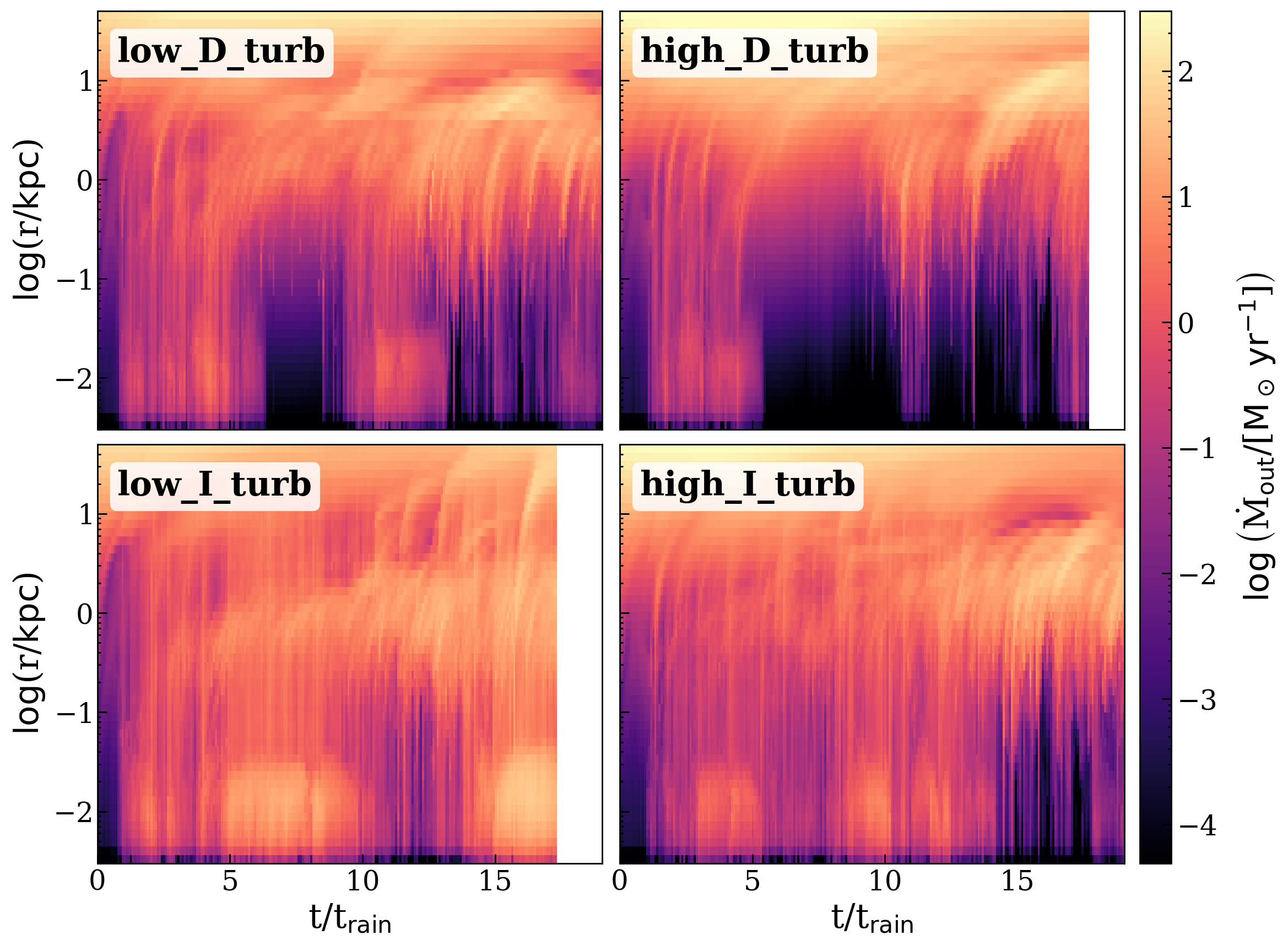}
\includegraphics[width=0.5\textwidth]{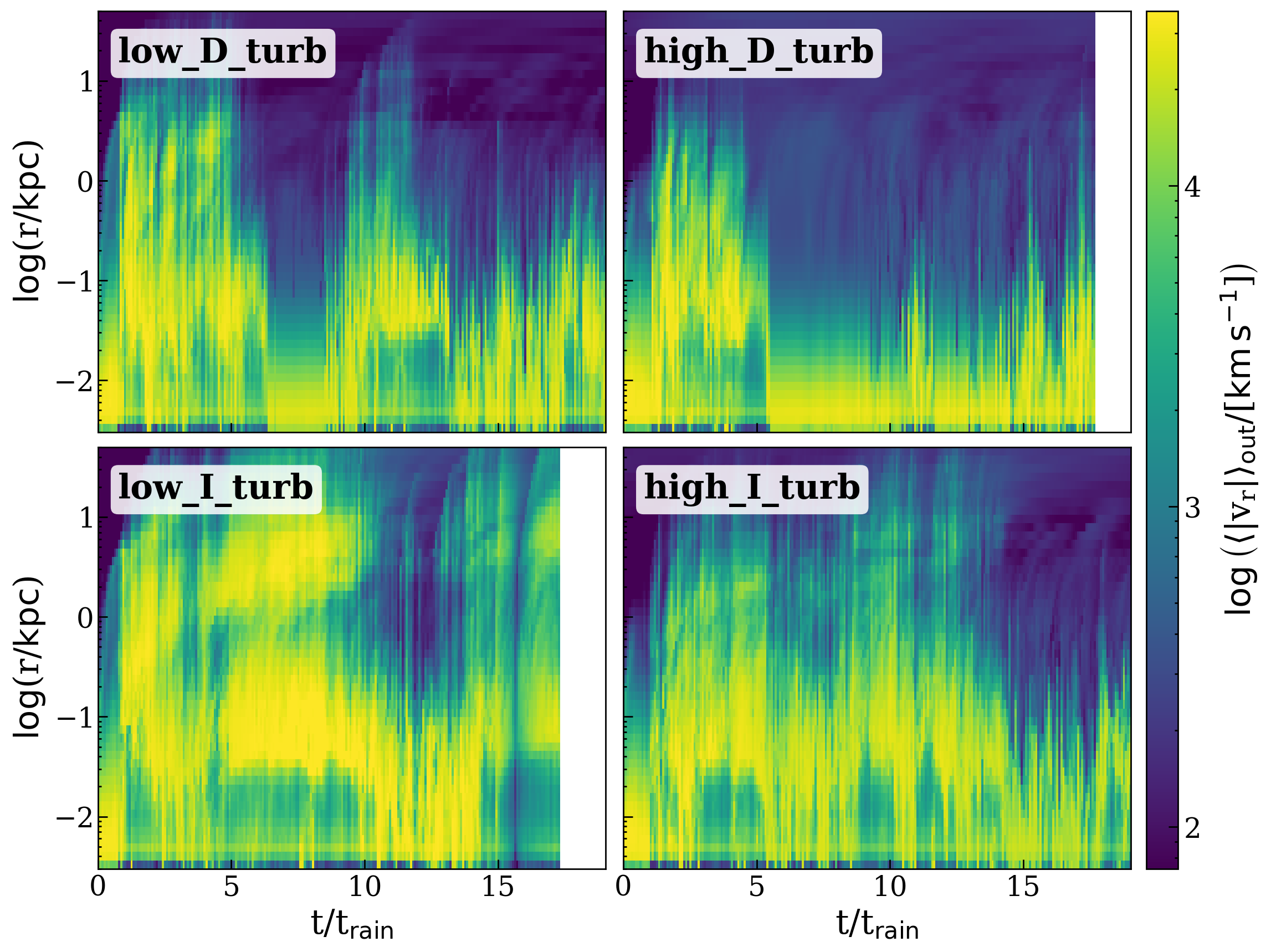}
\caption{Time--radius outflow mass-rate and radial-velocity maps for the whole gas. The total outflow largely follows the hot component but is broadened by entrained warm and cold material.}
\label{outflow_map}
\end{figure*}

The whole-gas outflow in Figure~\ref{outflow_map} is useful mainly as a consistency check. It reproduces the broad timing of the hot outflow shown in the main text, because the hot phase dominates the energetics of the launched gas, but it also includes the entrained warm and cold components. The resulting pattern is therefore physically complete but less clean as a weather-cycle diagnostic than the hot outflow alone.

\begin{figure*}[t]
\includegraphics[width=0.5\textwidth]{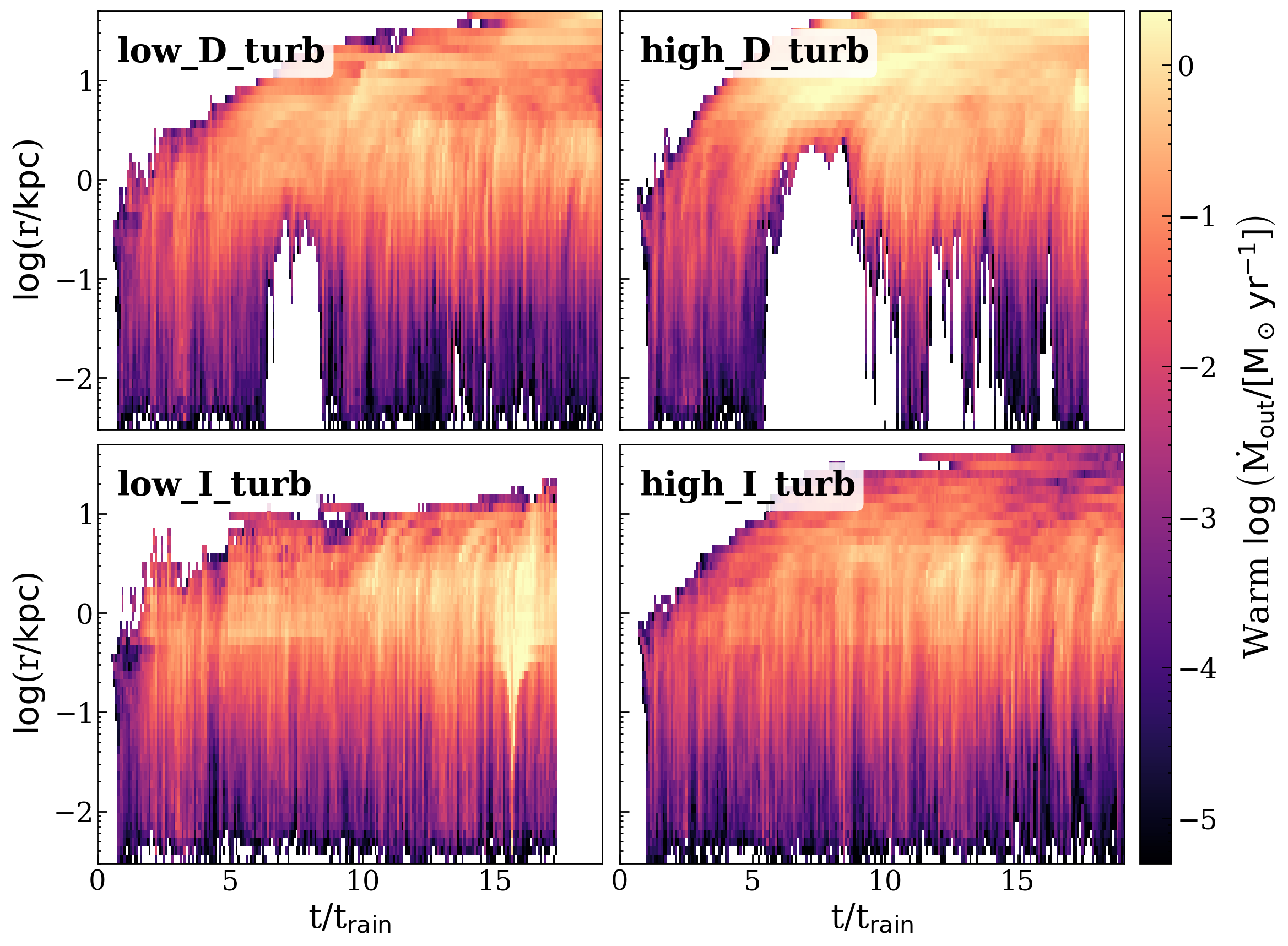}
\includegraphics[width=0.5\textwidth]{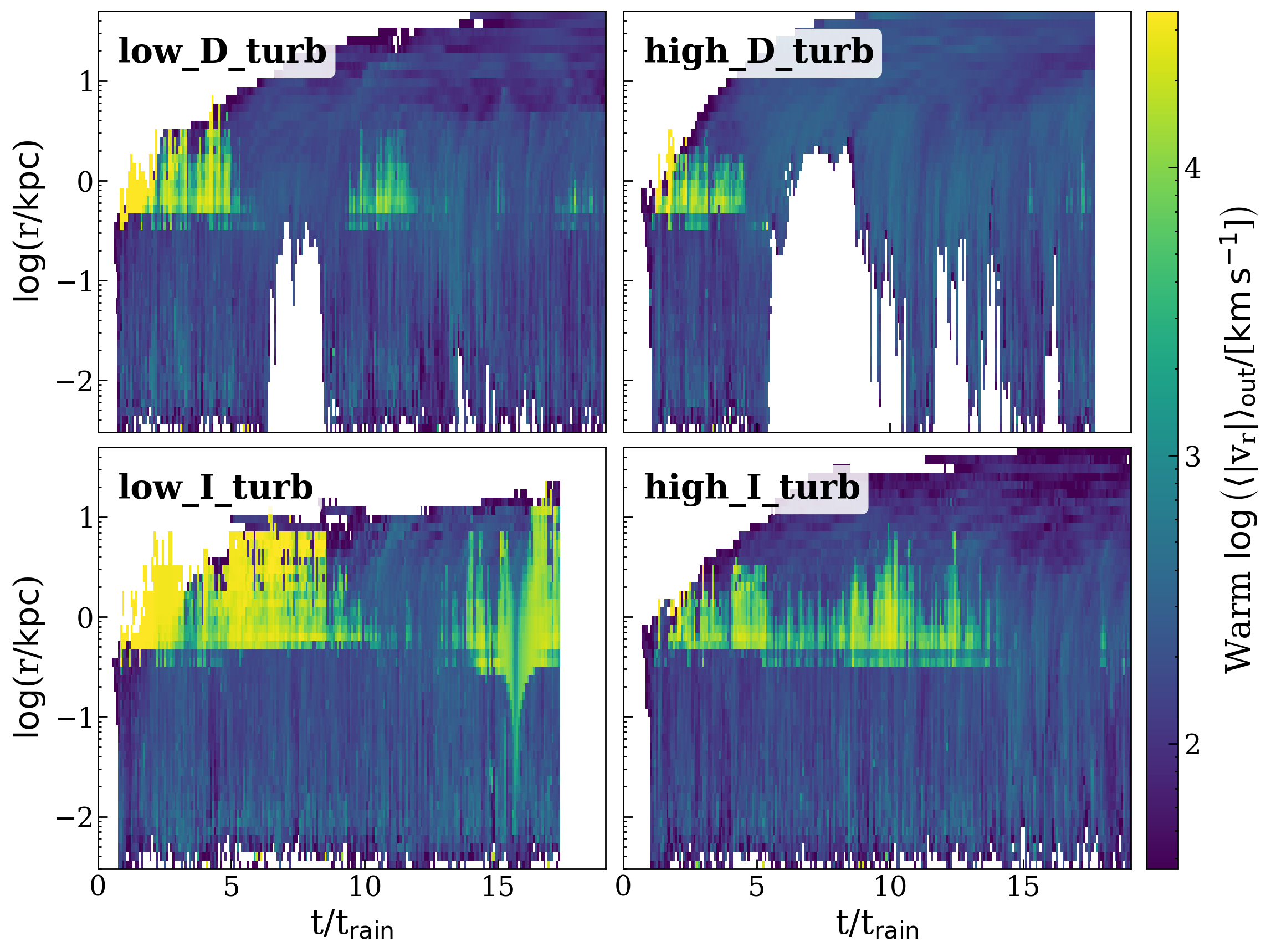}
\caption{Time--radius outflow mass-rate and radial-velocity maps for the warm phase. Warm outflow traces entrainment on intermediate scales and remains more extended and coherent in the IT suite than in the DT suite.}
\label{outflow_map_warm}
\end{figure*}

\begin{figure*}[t]
\includegraphics[width=0.5\textwidth]{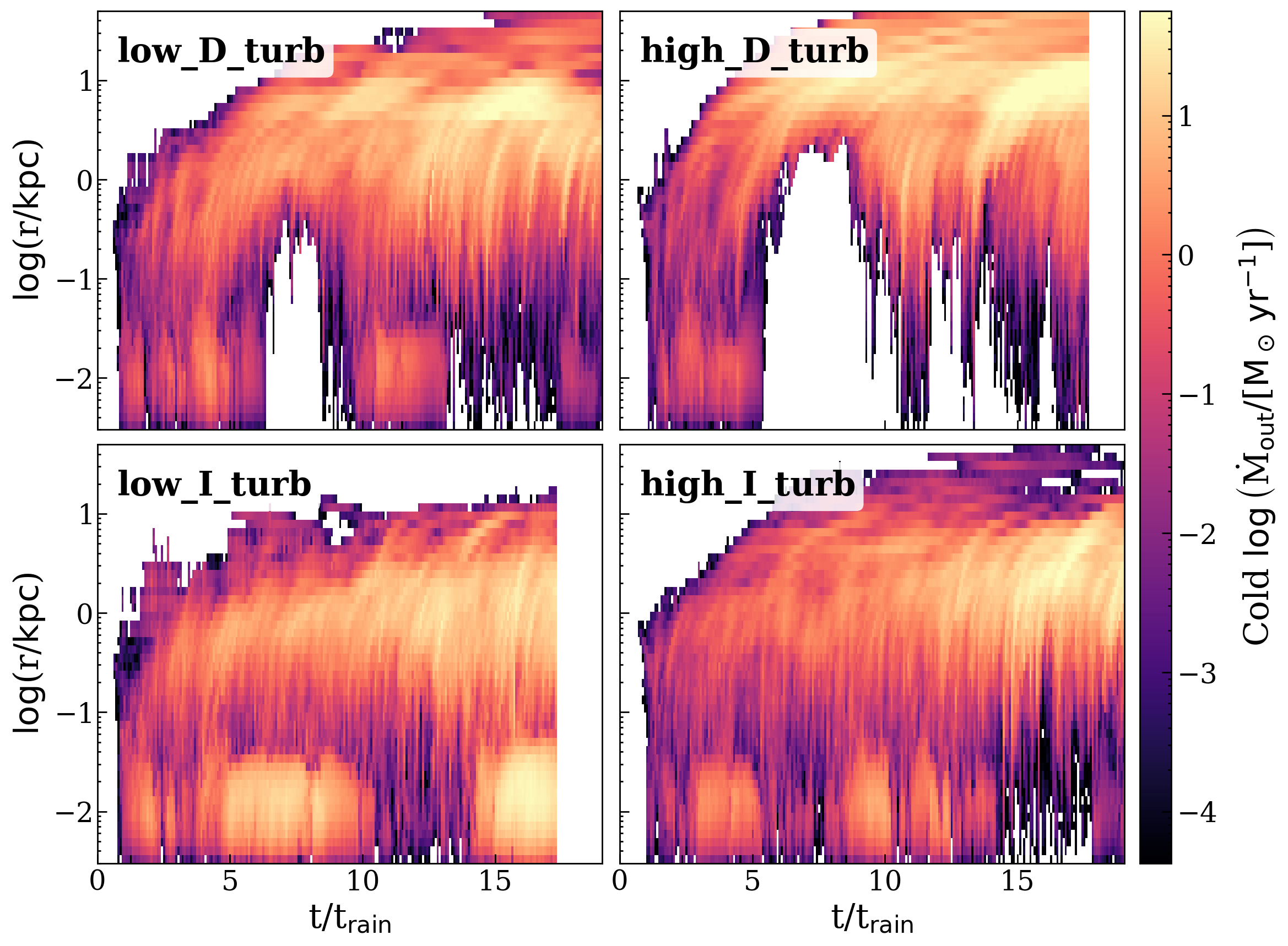}
\includegraphics[width=0.5\textwidth]{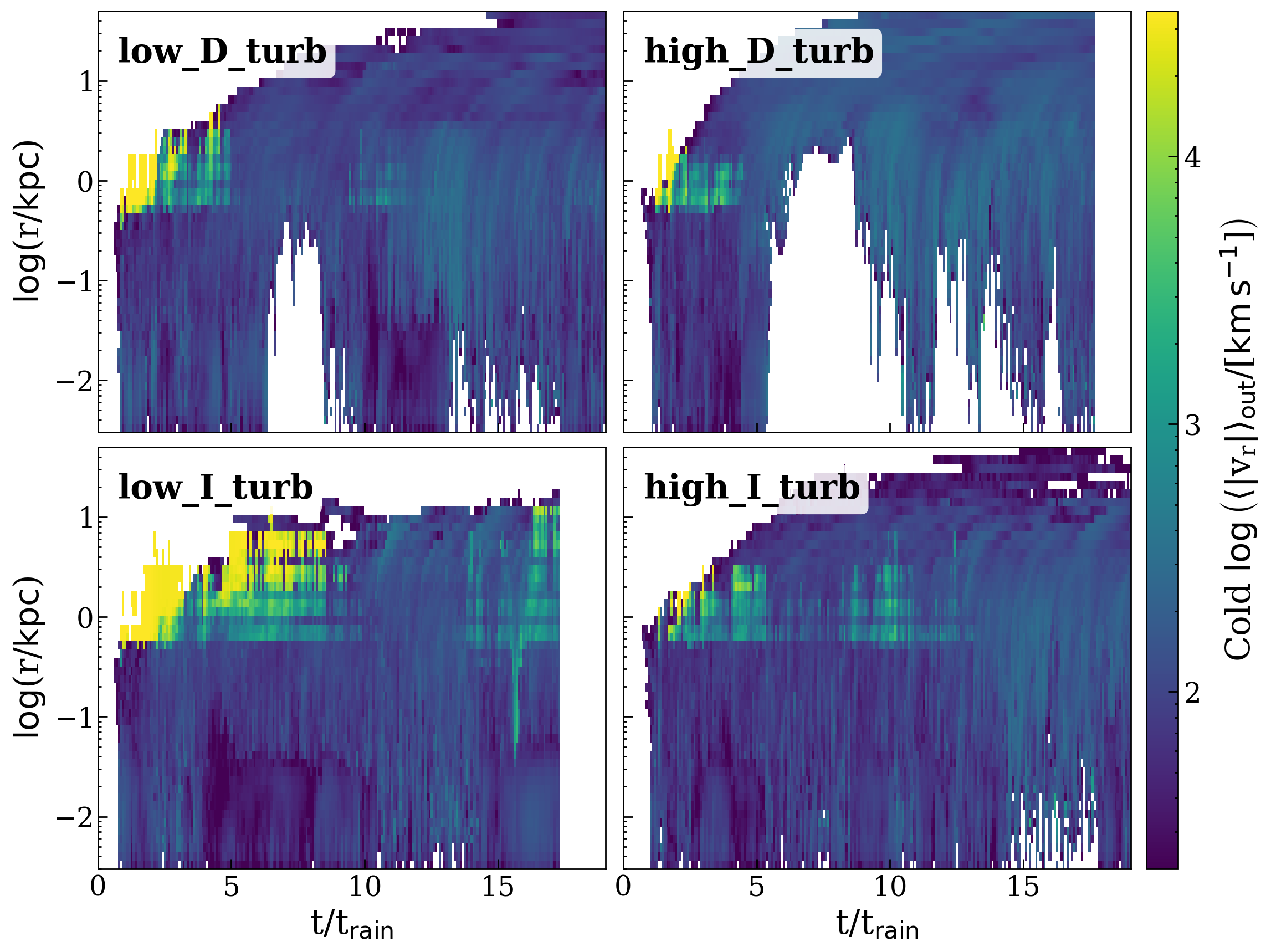}
\caption{Time--radius outflow mass-rate and radial-velocity maps for the cold phase. The cold outflow is the weakest and most intermittent component, confined to brief episodes and narrow radial ranges.}
\label{outflow_map_cold}
\end{figure*}

The warm and cold outflow maps show how entrainment propagates the nuclear feedback into the condensed phases. The warm outflow in Figure~\ref{outflow_map_warm} is strongest on intermediate scales, typically $r\sim0.3$--$10\,{\rm kpc}$, and is much more extended than the cold outflow. It is also visibly more coherent in the IT runs, where rainy phases provide more material to entrain, whereas the DT runs produce narrower and more episodic warm outflow structures. The cold outflow in Figure~\ref{outflow_map_cold} is the weakest and most intermittent component. It is confined to short episodes and narrow radial bands, indicating that direct lifting of the cold gas is not the dominant large-scale channel; most of the dynamically important feedback coupling occurs through the hot phase and, secondarily, through entrained warm gas.

\end{document}